\documentclass[twocolumn]{aastex631}

\usepackage{graphicx}	
\usepackage{amsmath}	
\usepackage{amssymb}	
\usepackage{xcolor}
\usepackage{cancel}
\usepackage{multirow}

\newcommand{\mcl}{M_{\rm cl}}
\newcommand{\msolt}{M$_\odot$~}
\newcommand{\mpcsqt}{M$_\odot$~pc$^{-2}$}
\newcommand{\pms}{$(P/m_\star)_0$}
\newcommand{\be}{\begin{equation}}
\newcommand{\ee}{\end{equation}}
\newcommand{\ie}{\emph{i.e.}, }
\newcommand{\eg}{\emph{e.g.}, }

\definecolor{color1}{HTML}{FF6C0C}

\newcommand{\rev}[1]{\textcolor{black}{#1}}

\received{XXX}
\revised{YYY}
\accepted{ZZZ}

\submitjournal{ApJ}

\shorttitle{Bursting Superbubbles}
\shortauthors{M. E. Orr et al.}

\usepackage{totcount}
\newtotcounter{citenum}
\def\oldcite{}
\let\oldcite=\bibcite
\def\bibcite{\stepcounter{citenum}\oldcite}

\begin{document}


\title{Bursting Bubbles: Feedback from Clustered SNe and the Trade-off Between Turbulence and Outflows}

\correspondingauthor{Matthew Orr}
\email{matt.orr@rutgers.edu}

\author[0000-0003-1053-3081]{Matthew E. Orr}
\affiliation{Department of Physics and Astronomy, Rutgers University, 136 Frelinghuysen Road, Piscataway, NJ 08854, USA}
\affiliation{Center for Computational Astrophysics, Flatiron Institute, 162 Fifth Avenue, New York, NY 10010, USA}
\affiliation{TAPIR, Mailcode 350-17, California Institute of Technology, Pasadena, CA 91125, USA}

\author[0000-0003-3806-8548]{Drummond B. Fielding}
\affiliation{Center for Computational Astrophysics, Flatiron Institute, 162 Fifth Avenue, New York, NY 10010, USA}

\author[0000-0003-4073-3236]{Christopher C. Hayward}
\affiliation{Center for Computational Astrophysics, Flatiron Institute, 162 Fifth Avenue, New York, NY 10010, USA}

\author[0000-0001-5817-5944]{Blakesley Burkhart}
\affiliation{Department of Physics and Astronomy, Rutgers University, 136 Frelinghuysen Road, Piscataway, NJ 08854, USA}
\affiliation{Center for Computational Astrophysics, Flatiron Institute, 162 Fifth Avenue, New York, NY 10010, USA}

\begin{abstract}
We present an analytic model for clustered supernovae (SNe) feedback in galaxy disks, incorporating the dynamical evolution of superbubbles formed from spatially overlapping SNe remnants.  We propose two realistic outcomes for the evolution of superbubbles in galactic disks: (1) the expansion velocity of the shock front falls below the turbulent velocity dispersion of the ISM in the galaxy disk, whereupon the superbubble stalls and fragments, depositing its momentum entirely within the galaxy disk, or (2) the superbubble grows in size to reach the gas scale height, breaking out of the galaxy disk and driving galactic outflows/fountains.  In either case, we find that superbubble breakup/breakout \emph{almost always} occurs before the last Type-II SN  ($\lesssim$40 Myr) in the recently formed star cluster, assuming a standard high-end IMF slope, and scalings between stellar lifetimes and masses. The threshold between these two cases implies a break in the effective strength of feedback \emph{in driving turbulence within galaxies}, and a resulting change in the scalings of, \emph{e.g.}, star formation rates with gas surface density (the Kennicutt-Schmidt relation) and the star formation efficiency in galaxy disks. 
\end{abstract}






\section{Introduction}
Star formation does not happen in a vacuum, diffuse though the interstellar medium may be.  In order to understand how the stellar components of galaxies grow, it is critical to understand both how young stars shape their environments, and how those environments shape (and guide) star formation (see \citealt{Krumholz2018c} \& \citealt{Tacconi2020}, for recent reviews on star cluster formation and the evolution of the star-forming ISM, respectively, in galaxies across cosmic time).  Broadly speaking in the context of star formation, feedback from young stars comes in two varieties: prompt feedback involving physical processes with little to no delay time between the formation of stars and their effects, like photoionization and stellar winds/jets, and delayed feedback in the form of supernovae (SNe).

Work has shown that the effects of prompt feedback are crucial in shaping star formation \emph{locally} within giant molecular clouds (GMCs).  That is, processes like protostellar heating, photo-ionizing radiation, or stellar winds, are critical to setting the formation efficiency and density profiles of star clusters, the shape \emph{and} peak of the stellar initial mass function (IMF), the distribution and phase of gas, and the lifetimes of GMCs/duration of star formation events themselves \citep{Murray2010, Dale2014, Guszejnov2016, Grudic2018a}.  Except for the extremely diffuse outskirts of galaxies, which may find themselves thermally supported against fragmentation and collapse \citep{Schaye2004}, these feedback processes/heating mechanisms are generally not the primary regulator of galaxy disks beyond the confines of GMCs \citep{Krumholz2009, Ostriker2010, Smith2020arx}. 

However, `delayed feedback' (\ie supernovae) acts on scales exceeding GMCs.  Core-collapse (Type II, as well as Ib \& Ic) SNe\footnote{Type Ia SNe are another channel of this delayed feedback.  But their delay time is both too long ($\gtrsim 0.5$~Gyr) and their relative occurrence too low ($\sim 1$ SN per 1000 \msolt formed for Ia SNe versus $\sim 1$ SN per 100 \msolt for all core-collapse types) to be relevant for models regulating the ISM on approximately a galaxy dynamical time \citep{Maoz2012}.} are the primary injector of momentum back into the ISM on the scale of the gas disk height, \ie hundreds of parsecs up to approximately a kiloparsec \citep{Ostriker2010, Faucher-Giguere2013}.  The momenta of SNe shockwaves appears easily coupled to gas in the ISM, with exception of high-density bound structures, driving (highly) supersonic turbulence  \citep{Kim2015b, Martizzi2015, Martizzi2016, Padoan2016, Iffrig2017, Martizzi2020}.
Furthermore, these stellar explosions are able to drive galactic winds, outflows and fountains \citep{Fielding2017, Fielding2018} especially if they are not deeply embedded in the dense birth cloud \citep{Iffrig2017, Seifried2018, Lucas2020} and if some previous event has cleared a channel out of the disk/cloud.

Galaxies have long been thought to be marginally stable against gravitational fragmentation and collapse \citep{Toomre1964, Goldreich1965, BinneyTremaine1987, Dekel2009}, with the global balance of turbulent energy/momentum in the ISM maintaining a rough equilibrium.  Feedback-regulated galactic star formation models have invoked the injection of feedback (both prompt \emph{and} delayed) as the key mechanism that maintains this balance \citep[\eg][]{Ostriker2010, Faucher-Giguere2013, Hayward2017}.  One way or another many galaxy disk models explicitly assume a threshold for fragmentation/collapse that can be quantified by some form of Toomre's Q-criterion \citep{Toomre1964}, \eg
\be \label{eq:Q}
\tilde Q_{\rm gas} = \frac{\sqrt{2} \sigma \Omega}{\pi G \Sigma_{\rm disk}}
\ee
which holds for a disk with a flat rotation curve\footnote{Up to an order unity pre-factor, this is nearly identical to the local Virial parameter when considering length scales of the gas disk height, as $\alpha_{\rm vir} \approx \sigma^2 H / GM \approx \sigma^2/G \Sigma H \approx \sigma \Omega / G \Sigma \approx \tilde Q$. And so, some form of $Q$ or $\alpha$ is inescapable.}, where $\sigma$ is a 1-D velocity dispersion (interchangeably $\sigma_R$, $\sigma_\phi$, or $\sigma_z$, if turbulence is isotropic) $\Omega \equiv v_c/R$ 
is the inverse dynamical time ($v_c$ being the local circular velocity, and $R$ the local galactocentric radius),  $G$ is the Newtonian gravitational constant, and $\Sigma_{\rm disk} = \Sigma_g + \Sigma_\star = \Sigma_g/\tilde f_g$ is the local disk surface density (neglecting the small contribution of dark matter near the disk plane) with $\tilde f_g$ being the local gas fraction. We use this definition for $\tilde Q_{\rm gas}$ (and $\tilde f_g$) throughout the paper.  For $\tilde Q_{\rm gas} > 1$, we expect gas disks to be stable with turbulent (or thermal) energies or local galactic shear exceeding the local gravitational forces.  With sub-unity values, however, we expect fragmentation and collapse to occur in gas disks.  Hence $\tilde Q_{\rm gas} = 1$ is the attractor state of gas in galaxies: sub-unity regions of gas fragment and collapse into stars, removing those regions from the galactic gas reservoir, and the turbulent or thermal support for stable gas slowly decays to the $\tilde Q_{\rm gas}=1$ threshold.  A large emphasis has thus been placed on understanding the gas turbulence/velocity dispersions \citep[\eg][]{Goldbaum2015,Krumholz2018,Orr2019, Orr2020}, as the evolution of $\sigma$ is on the timescale of a disk crossing time ($\approx H/\sigma \approx 1/\Omega$), whereas $\Sigma_{\rm disk}$ or $\Omega$ itself are tied to longer, cosmological timescales. These arguments, when considering warm atomic gas ($T \approx 10^4$~K, where $c_s \approx 11$~km/s), have also been invoked to explain the radial extent of star-forming disks \citep{Schaye2004, Stark2017, Orr2018} 

Significant work has attempted to calibrate the feedback momentum injected into the ISM per SN explosion. Early work investigated the balance of energy and momentum conservation, and the general nature of the astrophysical blastwaves \citep[see][for an early review on the subject]{Ostriker1988}.  However, focus shifted to quantifying the terminal momentum of the shockwaves expanding into the ISM \citep{Cioffi1988, Thornton1998, Martizzi2015}.  Galaxy simulations (and simulations of disk patches) which do not spatially resolve the full evolution of SN explosions, most importantly the early Sedov-Taylor stage, often use these calibrations to set the strength of feedback in their physics models (\eg see the careful SN treatment by \citealt{Kim2017} and \citealt{Hopkins2018:SN} in the {\scriptsize TIGRESS} and {\scriptsize FIRE-2} simulations, respectively, which inject a constant radial momentum per SN when the Sedov-Taylor stage is not adequately resolved).  By doing so, simulations generally `close the loop' of turbulence sourcing (feedback/gravitational migration) and dissipation\footnote{Ignoring the effects of (non-physical) numerical dissipation in the simulations.} (via hydrodynamic interactions) in their galaxies, and the rate of star formation comes into an equilibrium to maintain the appropriate momentum that produces $\tilde Q_{\rm gas} \approx 1$.

A nuance, implicitly accounted for in galaxy simulations that adequately resolve a multiphase ISM with a realistic star formation prescription (\eg the cosmological zoom-ins of the {\scriptsize FIRE} suite \citealt{Hopkins2014, Hopkins2018:fire}, or the tall-box {\scriptsize TIGRESS} simulations of \citealt{Kim2017}), is the effect of spatially clustered SNe.  Though Type-Ia SNe loosely occur in regions with star formation in the past several Gyr, the correlation is hardly more than a radial gradient in their rates, owing to the long/wide delay time distribution \citep{Tsvetkov2017}; whereas core-collapse SNe are highly correlated with star-forming regions, as only massive stars with high peculiar velocities relative to their birth clouds can escape to appreciable distances in the short time window ($\sim$40 Myr) before detonating.  And so, for the star cluster environment in which most stars form, SNe occur closely spaced in time and space, affecting the ability of individual SNe to effectively couple their terminal momentum into the ISM or drive galactic outflows/fountains \citep{Fielding2018, Martizzi2020}.

\rev{Although theoretical work investigating the nature of the overlapping SN shock fronts is quite mature (\eg \citealt{Weaver1977, Tomisaka1986, Maclow1988, Maclow1989, Koo1992}), recent simulation work has significantly altered our understanding of the dynamical state/evolution of these superbubbles \citep{Kim2016, Fielding2018, Oku2022}.  Work focusing on cooling, conduction and stratified media found that these bubbles likely evolved while entirely remaining in an energy-conserving state \citep{Tomisaka1986, Maclow1989, Koo1992}.  However, more modern 3-D simulations which include turbulent mixing with the cold dense ISM, as well as the effects of other ``pre-processing" feedback like stellar winds and photoionization have concluded that superbbubles quickly radiate thermal energy at the turbulent shock front, entering a momentum-driven regime, with a roughly constant momentum coupling per supernova \citep{Kim2016, Fielding2018, Oku2022}. This, in spite of the panoply of physical processes at play allows us to consider a very simple model for the effects of clustered SNe on the disk scale in galaxies.}

In this paper, we will explore the effects of (spatially and temporally) clustered core-collapse SNe feedback, in the form of superbubbles expanding into a galactic gas disk, on the \emph{effective} strength of feedback and galactic outflows/fountains.  In \S~\ref{sec:model}, we will develop a simple model for superbubbles, derived from assumptions regarding star cluster formation efficiency, the IMF/stellar lifetimes, and invoking marginal Toomre stability.  \rev{We specifically build our model using scalings found in the star cluster formation simulations of \citet{Grudic2018}, and the superbubble simulations of \citet{Fielding2018}, so as to interpret their results in a galaxy scale context.} The following \S~\ref{sec:outcomes} will explore the evolutionary outcomes of these bubbles, as either stalling/fragmenting remnants or producing galactic outflows. Sections \ref{sec:turbscale}--\ref{sec:outflows} investigate the model's predictions regarding turbulence driving, the Kennicutt-Schmidt relation, and the properties of outflows. The reader will find a broader discussion of the model in the context of star formation/galaxy evolution literature in \S~\ref{sec:discussion}. Lastly, we summarize our results and conclude in \S~\ref{sec:conclusion}. 
\section{Superbubble Model}\label{sec:model}

\begin{figure*}
\centering
	\includegraphics[width=\textwidth]{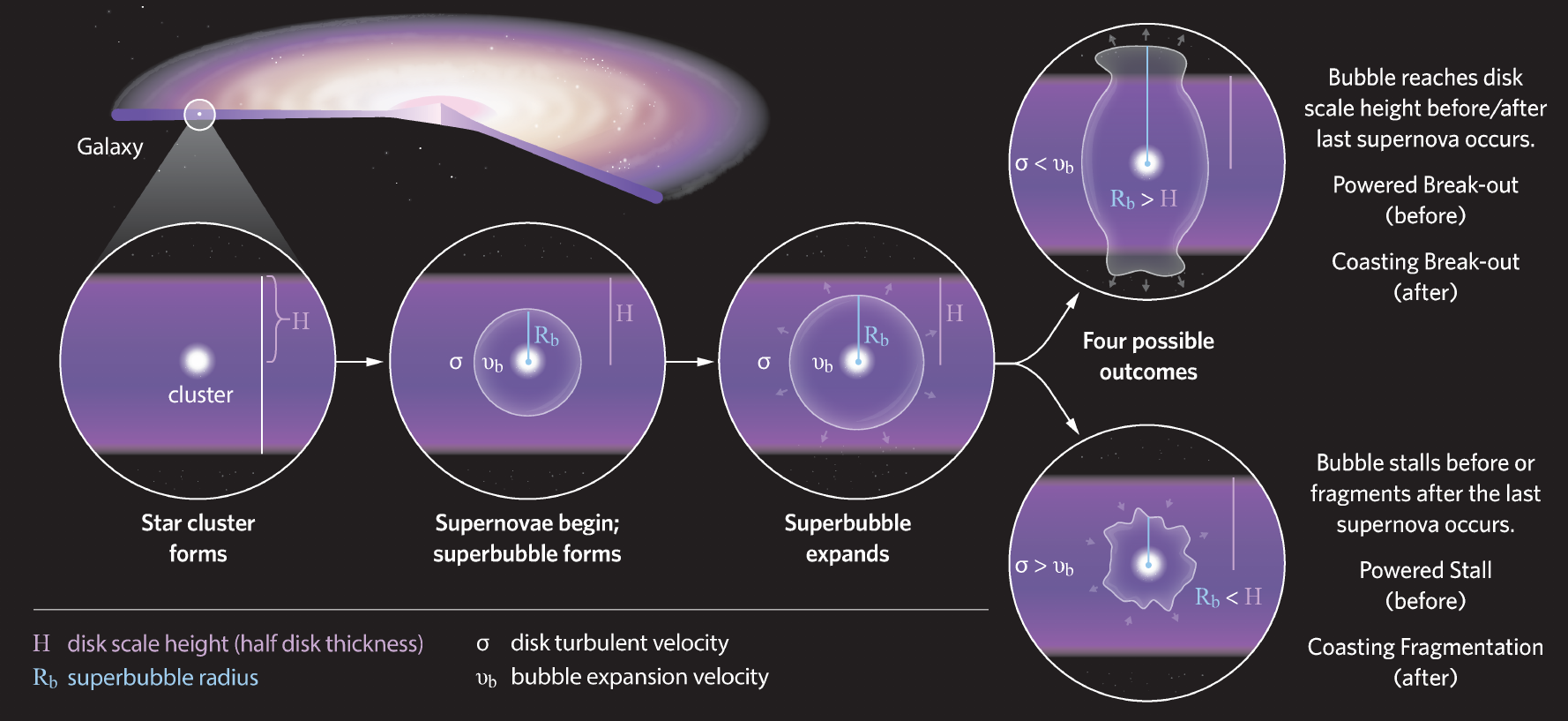}
    \caption{Schematic of the superbubble evolution and outcomes. \emph{Insets, left to right:} A young star cluster forms in the ISM; After a period corresponding to the lifetime of the most massive star formed, a shock front forms from the overlapping SNe, and begins to expand into the ISM. Possible outcomes from the evolution of the SNe remnant include: \emph{(top right)} expansion until the bubble reaches the gas disk scale height, whereupon the bubble breaks out and drives galactic outflow/fountain phenomena, either before (\textbf{PBO case}) or after (\textbf{CBO case}) the last SN occurs, or \emph{(bottom right)} stall and fragmentation of the bubble as it comes into pressure equilibrium ($v_b \leq \sigma$) with the ISM before/after (\textbf{PS/CF cases}) the last SN occurs.}
    \label{fig:cartoon}
\end{figure*}

Following the formation of a star cluster of mass $\mcl$ in a GMC, we expect that after a short period $t_d$ ($\sim 3$ Myr), corresponding to the lifetime of the most massive star formed in that cluster, core-collapse SNe (hereafter referred to simply as SNe) begin to occur.  For the purposes of this model, we take the formation of the star cluster to be instantaneous, as star formation generally appears to stop in clouds before the time that the first SN occurs, with observed and simulated star clusters having age spreads of 1-3 Myr \citep{Murray2011, Grudic2018}. \rev{Moreover, we implicitly account for the various prompt feedback processes like radiation pressure and stellar winds in this manner: that they primarily shape the IMF and set the local efficiency of star cluster formation.} Given that the number of supernovae from the cluster $N_{\rm SNe} \approx \mcl/100$ M$_\odot$\footnote{The 100 M$_\odot$ value is derived from the fraction of stars $>$8 M$_\odot$ (SN cutoff mass, \citealt{Smartt2009}) formed in a young star cluster, assuming a \citet{Kroupa2002} IMF.} occur over a relatively short period of $t_{\rm SNe} {\sim}40$ Myr, the supernovae remnants temporally and spatially overlap forming a cavity/void in the ISM that expands as a superbubble (provided that the time between supernovae is shorter than the time for an individual to come into pressure equilibrium with the surrounding ISM, see Eq.~6 of \citealt{Fielding2018}).  The expansion of this superbubble continues until the shock front either comes into pressure equilibrium with the surrounding ISM, or it will break out of the gas disk and drive a galactic fountain/outflow (see \S~\ref{sec:outcomes}).  Figure~\ref{fig:cartoon} illustrates the general model, and enumerates the outcomes of the superbubble evolution in the ISM.

\begin{table}\caption{Summary of variables used in this paper}\label{table:variables}
\centering
\begin{tabular}{ll}
\hline
 Symbol & Definition  \\
 \hline
$\tilde Q _{\rm gas}$		& Modified Toomre-Q gas stability parameter \\
$\sigma$				& Turbulent gas velocity dispersion (1-D) \\
$\Omega$				& Local orbital dynamical time \\
$H$ 					& Gas scale height \\
$G$ 					& Newtonian gravitational constant  \\
$M_{\rm cl}$			& Star cluster mass \\
$\epsilon_{\rm int}$ 	& Integrated star cluster formation efficiency \\
$\Sigma_{\rm crit}$ 		& \rev{Gas surface density of maximum star cluster}\\
			 		& \rev{formation efficiency}\\
$\Sigma_{\rm disk}$ 		& Local disk surface density \\
$\Sigma_g$ 			& Local gas surface density \\
$\tilde f_g$ 			& Local fraction of disk mass in gas \\
$\bar \rho_g$			& Disk mid-plane gas volume mass density \\ 
$R_b$				& Superbubble shock front radius \\
$v_b$				& Superbubble shock front velocity \\
$t_d$ 				& Delay timescale until first supernova   \\
$t_{\rm SNe}$ 			& Supernova feedback duration   \\
$\alpha$ 				& Power law slope of Type-II SNe delay\\
   					& time distribution \\
$(P/m_\star)_0$ 		& Normalized feedback momentum    \\
   					& per mass of stars formed   \\
$\dot\Sigma_\star$ 		& Star formation rate surface density  \\
$t_{\rm BO}$			& Time of superbubble break-out \\
\rev{$t_{\rm frag}$}			& \rev{Time of superbubble stall/fragmentation} \\

 \hline
\end{tabular}
\end{table}

\subsection{Star Cluster Formation Model}\label{sec:SCFM}
Following the fits to star cluster formation efficiency in simulations by \citet{Grudic2018}, we will assume that the young star cluster forms with an integrated efficiency proportional to the gas surface density of the ISM, \ie $\epsilon_{\rm int} \approx \Sigma_g/\Sigma_{\rm crit}$, such that
\be \label{eq:mcl}
\mcl = \epsilon_{\rm int} M_g \approx \pi H^2 \Sigma_g^2/\Sigma_{\rm crit} \; ,
\ee
assuming that $M_g \approx \pi H^2 \Sigma_g$ (the Toomre mass of self-gravitating clumps), and taking the ``critical surface density'' from \citet{Grudic2018} to be $\Sigma_{\rm crit} = 2800$ \mpcsqt.  As for most galactic disk conditions $\Sigma_g \ll \Sigma_{\rm crit}$, formation efficiencies are fairly small, and we can safely neglect the subtraction of this gas from the ISM for the subsequent evolution of our model superbubbles (\ie we do not need to account for a term $\sim$$(1 - \Sigma_g/\Sigma_{\rm crit})\approx 0.97-0.99$ throughout). Taking $H \approx \sigma/\Omega = \sigma R/ v_c$, the typical scale heights we would expect in a local disk galaxy are on the order 10 km/s $\times$ 5  kpc / 200 km/s $\sim$ 250 pc (\ie a few hundred parsecs).

Furthermore, by using an integrated star formation efficiency set by $\Sigma_{\rm crit}$,  we are implicitly including the effects of early/prompt feedback (\eg photo-ionizing radiation or stellar winds) as those physics appear key in setting the \emph{local} star formation efficiency, in terms of the direct conversion between dense gas mass \emph{within} a GMC and subsequent young star clusters \citep{Dale2014, Crocker2018, Grudic2018, Grudic2019a, Kruijssen2019, Li2019, Grudic2020, Kim2020}.

\subsection{SNe Remnant Evolution in \rev{a Simplified ISM}}\label{sec:remnantevolution}

\begin{table}\caption{Superbubble model outcome case structure} \label{table:cases}
\centering
\begin{tabular}{l|c}
\hline
Case Description & \hspace{-3em} {\begin{tabular}[c]{@{}c@{}} Outcome \\ \emph{Timing Relative to $t_{\rm SNe}$}\end{tabular}} \\ \hline
 \hspace{-3em}{\begin{tabular}[c]{@{}l@{}}\textbf{PBO}\\ (``Powered Break-out")\end{tabular}} & \hspace{-3em}{\begin{tabular}[c]{@{}c@{}}Bubble breaks out\\ $t_{\rm BO} < t_{\rm SNe}$\end{tabular}} \\   \hline
    \hspace{-3em}{\begin{tabular}[c]{@{}l@{}}\textbf{CBO}\\ (``Coasting Break-out")\end{tabular}} & \hspace{-3em}{\begin{tabular}[c]{@{}c@{}}Bubble breaks out\\ $t_{\rm BO} > t_{\rm SNe}$\end{tabular}} \\   \hline
   \hspace{-3em} {\begin{tabular}[c]{@{}l@{}}\textbf{PS}\\ (``Powered Stall")\end{tabular}} & \hspace{-3em}{\begin{tabular}[c]{@{}c@{}}Bubble Stalls\\ $t_{\rm frag} < t_{\rm SNe}$\end{tabular}} \\   \hline
 \hspace{-3em}   {\begin{tabular}[c]{@{}l@{}}\textbf{CF}\\ (``Coasting Fragmentation")\end{tabular}} & \hspace{-3em}{\begin{tabular}[c]{@{}c@{}}Bubble Fragments\\ $t_{\rm frag} > t_{\rm SNe}$\end{tabular}} \\    \hline
\end{tabular}
\end{table}

\begin{figure}
	\includegraphics[width=0.97\columnwidth]{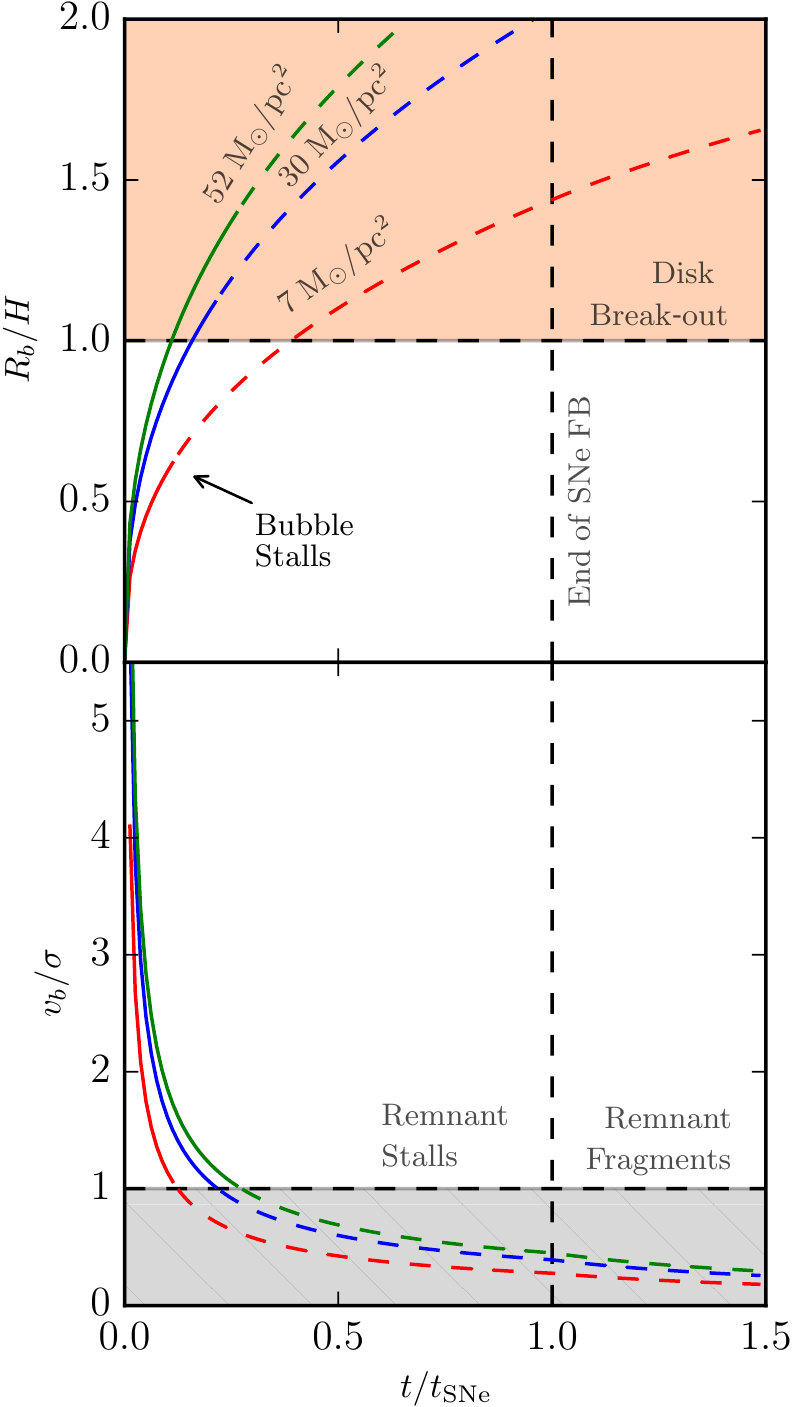}
    \caption{Evolution of superbubble radius (top panel, following Eq.~\ref{eq:rb}) and shock-front velocity (bottom panel, following Eq.~\ref{eq:vb}), assuming $\tilde Q_{\rm gas}=1$, $H=300$ pc and $\Omega = 50$ Gyr$^{-1}$ (thus $\sigma \approx H\Omega \approx 15$ km/s), for three disk gas fractions $\tilde f_g = 0.1, 0.4, 0.7$ (colored lines).  Change in line style represents the point at which the shock-front velocity falls below the gas disk velocity dispersion $\sigma$ (bottom panel $v_b/\sigma < 1$, grey shaded region), resulting in shock-front stalling/fragmentation (\emph{if} the shock-front has not already reached the gas disk scale height and broken out). We see that for the two most gas-rich cases ($\tilde f_g = 0.4, 0.7$) the remnants reach the disk scale height before stalling/fragmenting (top panel, disk break-out occurring in the orange shaded region) and thus \rev{represent superbubbles that successfully break out of the ISM}.  Perhaps counter-intuitively, the more gas rich the disk, the easier it is for superbubble remnants to break out.}
    \label{fig:evolve}
\end{figure}
\begin{table*}\caption{Fiducial Model Parameters}\label{table:fidmodel}
\begin{tabular}{llll}
\hline
Parameter & Quantity & Fiducial Value & Reference \\
 \hline 

Critical $\Sigma_g$ for Star Formation 	& $\Sigma_{\rm crit}$	& 2800 \mpcsqt	& \citet{Grudic2018} \\ 
Efficiency Saturation & & & \\
Normalized Feedback Strength			& $(P/m_\star)_0$				& 3000 km/s  & \citet{Martizzi2015} \\ 
SNe Feedback Duration		& $t_{\rm SNe}$ 					& 40 Myr & \citet{Raiteri1996} \\ 
Power law slope of Type-II 	&$\alpha$ 				& 0.46 & Appendix A in \citet{Orr2019} \\
SNe delay time distribution 	&						& & \\

 \hline
\end{tabular}
\end{table*}

We are considering a Toomre regime where GMCs are coupled to the ISM, rather than a decoupled GMC phase. And so, we assume that the ``bubble", \ie expanding overlapping SNe remnants, does not first have to break out of any overdense region and just consider it to be embedded in the ISM of a gas disk with a mean mass-density of $\bar \rho_g = \Sigma_g/2H$. For simplicity, we will assume locally a slab geometry\footnote{As this simple model neglects the gravitational work involved in lifting the expanding shell from the disk midplane, the particular choice in vertical density gradient is unimportant for our calculations.  Other analytic studies have attempted to model the particular geometry of evolving superbubbles \citep[\eg][]{Baumgartner2013}.}.  

\rev{We consider the superbubble formed from the overlapping SN remnants to evolve purely in a momentum-conserving regime.  Simulations of expanding superbubbles in a realistic ISM by \citet{Kim2016}, \citet{Fielding2017}, \citet{El-Badry2019}, and \citet{Oku2022} that included cooling processes, have shown that mixing of cold material from subsumed clouds in a turbulent medium rapidly causes the hot interiors of bubbles to radiate away their thermal energy and enter the momentum-conserving phase.  Moreover, that this transition to being momentum-driven happens shortly after the shock front of the superbubble forms, on the order of $\sim 10^4$ yr ($\ll t_{\rm SNe}$; see \citealt{Kim2016}, \citet{El-Badry2019}, and \citealt{Oku2022}, especially).}
The momentum contained in the shock front at radius $R_b$ of that bubble, having swept up the mass of gas within that radius, is,
\be \label{eq:pbub}
P_b = \frac{4}{3}\pi R_b^3 \bar\rho_g \frac{{\rm d}R_b}{{\rm d}t} \; ,
\ee
where ${\rm d}R_b/{\rm d}t \equiv v_b$ is the expansion velocity of the superbubble.

If this bubble is driven by the momentum injection from the SNe of the central star cluster, then the momentum of the bubble must be balanced with the cumulative momentum from the SNe that have occurred up until the current time.  This will involve the fact that the SNe are distributed in time.  A simple estimate for the time delay distribution of core-collapse SNe can be derived by convolving the high-end mass slope of the IMF ($dN_\star/dM_\star \propto M_\star^{-2.35}$) with an estimate for high-mass stellar lifetimes ($t_\star \propto M_\star/L_\star$, with $L_\star \propto M_\star^{3.5}$), yielding a fairly shallow (but non-zero) power-law for the SNe rate of d$N_{\rm SN}/$d$t \propto t^{-0.46}$ (see Appendix A of \citealt{Orr2019} for a more detailed discussion; throughout we will assume that the power-law slope $\alpha = 0.46$). We show a comparison of this delay time distribution to results from {\scriptsize STARBURST99} \citep{Leitherer1999, Leitherer2014} in Appendix \ref{appendix:SNrate}.  It is important that we make this choice, as the delay time distribution directly affects the cumulative momentum injected into the superbubble remnant in time (up until all the SNe have occurred).  \rev{As well, \citet{Kim2016} and \citet{Oku2022} showed in their suites of superbubble simulations that the final momentum coupled into the shock front per supernova was roughly constant.}
Generally speaking, we can write the cumulative momentum injected by SNe into the superbubble as,
\be
P_{\rm SNe}(t) = \mcl \left( \frac{P}{m_\star} \right)_0 \rev{\frac{(1-\alpha)}{t^{1-\alpha}_{\rm SNe}}} \int^t_0 t'^{-\alpha} {\rm d}t' \; ,
\ee
where $(P/m_\star)_0$ is the fiducial momentum injected by a single supernova ($\approx 3000$ km/s, \citealt{Martizzi2015}\rev{, also \citealt{Iffrig2015, Kim2015b}}), normalized per 100 M$_\odot$, and the quantity $\rev{\frac{(1-\alpha)}{{t^{1-\alpha}_{\rm SNe}}}} \int^t_0 t'^{-\alpha} {\rm d}t'$ representing the normalized fraction of SNe to have occurred by the time $t$ (\emph{i.e.}, the quantity ranges from zero to one from the time of the first SN to the time of the last at $t_{\rm SNe}$). We neglect the short delay time term $t_d$ in the integral, \ie $\int_0^t (t'+\cancelto{0}{t_d})^{-\alpha} dt'$, to make the following analysis analytically tractable. Though it ``front loads" the feedback momentum slightly ($\sim$20\%) in time, by conserving the amount of total feedback momentum the results are not qualitatively altered.  Integrating this yields
\be \label{eq:psne}
P_{\rm SNe}(t) = \mcl \left( \frac{P}{m_\star} \right)_0 
\begin{cases}
\left( \frac{ t }{t_{\rm SNe}} \right)^{1-\alpha} , &  0 < t < t_{\rm SNe} \\
1 , & t > t_{\rm SNe} \; ,
\end{cases}
\ee
which, despite our choice in neglecting $t_d$, does not have any discontinuity at $t=0$.
And so, momentum conservation requires $P_b = P_{\rm SNe}$ at all times before the bubble breaks out of the disk (or breaks up).  Hence,
\be
\frac{2}{3}\pi R_b^3 \frac{\Sigma_g}{H} \frac{{\rm d}R_b}{{\rm d}t} = \mcl \left( \frac{P}{m_\star} \right)_0 \begin{cases}
\left( \frac{ t }{t_{\rm SNe}} \right)^{1-\alpha} , &  0 < t < t_{\rm SNe} \\
1 , & t > t_{\rm SNe} \; .
\end{cases}
\ee
Integrating this, and substituting in our assumption regarding cluster formation efficiency (Eq.~\ref{eq:mcl}), yields a relation for the radius of the bubble in time:

\be \label{eq:rb}
  R_b = H \left[ 6 \frac{\Sigma_g}{\Sigma_{\rm crit}} \frac{(P/m_\star)_0}{H/t_{\rm SNe}}  \right]^{\frac{1}{4}}
  \begin{cases}
  \!\begin{aligned}
       &\frac{1}{(2-\alpha)^{1/4}} \left( \frac{t}{t_{\rm SNe}}\right)^{\frac{2-\alpha}{4}}\\
       & \qquad \qquad 0 < t < t_{\rm SNe}
    \end{aligned} & \\
    \!\begin{aligned}
       &\left[\frac{t}{t_{\rm SNe}} -\left(\frac{1-\alpha}{2-\alpha}\right) \right]^{\frac{1}{4}}\\
       & \qquad \qquad t > t_{\rm SNe}
    \end{aligned}           
  \end{cases}
\ee
and the velocity $v_b$ ($\equiv {\rm d}R_b/{\rm d}t$) of the shock-front is,
 \be \label{eq:vb}
  v_b =\frac{H/t_{\rm SNe}}{4}\left[ 6 \frac{\Sigma_g}{\Sigma_{\rm crit}} \frac{(P/m_\star)_0}{H/t_{\rm SNe}}  \right]^{\frac{1}{4}}
  \begin{cases}
  
  \!\begin{aligned}
       &(2-\alpha)^{\frac{3}{4}} \left( \frac{t}{t_{\rm SNe}}\right)^{-\frac{(2+\alpha)}{4}}\\
       & \qquad \qquad 0 < t < t_{\rm SNe}
    \end{aligned} & \\
    \!\begin{aligned}
       & \left[\frac{t}{t_{\rm SNe}} -\left(\frac{1-\alpha}{2-\alpha}\right) \right]^{-\frac{3}{4}}\\
       & \qquad \qquad t > t_{\rm SNe}
    \end{aligned}           
    
  \end{cases}
\ee
%
We see that the velocity of the shell is monotonically falling with time, $v_b \propto t^{-(2+\alpha)/4}\approx t^{-0.62}$, even before the SNe stop occurring in the star cluster.  Further, the velocity of the shell falls more rapidly in the coasting phase, but not dramatically so, with $v_b \propto t^{-0.75}$.

Figure~\ref{fig:evolve} shows the evolution of three superbubble remnants (with differing gas fractions) following Eqs.~\ref{eq:rb} \& \ref{eq:vb}, for a marginally stable $\tilde Q_{\rm gas}=1$ disk, assuming $H=300$~pc and a dynamical time $\Omega = 50$ Gyr$^{-1}$ (such that $\sigma \approx H\Omega \approx 15$ km/s). The least gas rich case ($\tilde f_g = 0.1$) comes into pressure equilibrium, $v_b = \sigma$, before reaching the gas scale height, and subsequently stalls.  The other, more gas-rich cases reach the scale height before the shock-front slows to the local gas velocity dispersion.

\rev{Throughout, we will make the assumption that the Toomre-patch the superbubble evolves in is marginally stable $\tilde Q_{\rm gas}=1$.  This assumption is theoretically well-motivated on kiloparsec scales in disk galaxies, in feedback-regulation frameworks as well as disk models invoking hydrostatic balance \citep{Kim2007, Ostriker2011, Faucher-Giguere2013}, and seen in cosmological zoom-in simulations of spiral galaxies at late times \citep{Orr2020}.  Marginal Toomre-stability has also been confirmed in large samples of local star-forming spiral galaxies \citep{Leroy2008, Romeo2017}.  Doing so will transform the outcomes from being dependent on local velocity dispersion and gas surface density, to dependence on local gas fraction and (inverse) dynamical time.}


\section{Superbubble Outcomes in Disk Environments}\label{sec:outcomes}
\begin{figure}
\centering
	\includegraphics[width=0.97\columnwidth]{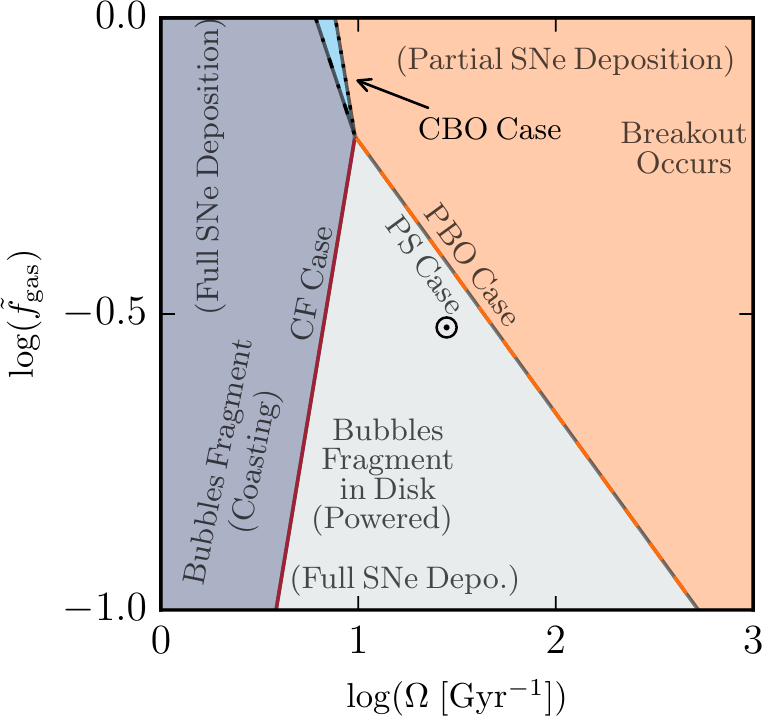}
    \caption{Gas fraction--dynamical time phase space of remnant outcomes, following Eqs.~\ref{eq:case0III}, \ref{eq:tbotsn}, \ref{eq:case23} \& \ref{eq:case12}.  Solid orange and light-blue regions denote cases where remnant successfully breaks out of the disk, and light-grey and blue-grey regions correspond to cases where the remnant fragments in the ISM.  Sun symbol indicates Solar Circle conditions ($\Sigma_\star \approx 35$ M$_\odot$, $\Sigma_{\rm gas} \approx 15$ M$_\odot$, and $\Omega \approx 35$ Gyr$^{-1}$; \citealt{McKee2015}). As \textbf{PBO/PS cases} occur before $t_{\rm SNe}$, these cases supersede the outcomes of \textbf{CBO/CF cases}, which occur in the unpowered coasting phase of remnant evolution after $t_{\rm SNe}$.  Under practical conditions, superbubbles almost always appear to either break out of the disk, or their expansion stalls within it, \emph{before} $t_{\rm SNe}$.  Only a small wedge of parameter space allows for break-out \emph{after} $t_{\rm SNe}$.  This implies that \emph{almost all} fountains/outflows are \emph{powered}, at least initially.}
    \label{fig:case}
\end{figure}
The ultimate fate of the evolution of these superbubbles in the ISM can be broken down into four cases, relating to whether the bubbles break out of the disk (at a time $t_{\rm BO}$), and if the central star cluster is still producing SNe.  The latter condition simply relates to the duration of the superbubble expansion, and whether or not the central engine (the young star cluster) continues to produce SNe (\ie $t<t_{\rm SNe}$), actively powering the expansion of the remnant.

Breakout itself is not guaranteed, as the superbubbles may not maintain coherence in expanding to reach the gas disk scale height.  \citet{Fielding2018} showed in simulations that fragmentation/stalling of the shock-front from overlapping SNe occurs when the expanding bubble comes into pressure equilibrium with the surrounding ISM (for a turbulent ISM, $P_{\rm ISM} \sim \bar \rho_g \sigma^2$). \rev{At this point a fragmentation of the entire superbubble volume occurs, as the ram pressure of the bubble no longer exceeds the turbulent ISM ram pressure.} This condition implies that the superbubbles stall and fragment when $v_b \approx \sigma$, demarcating the difference between cases where the remnant successfully and unsuccessfully reaches the disk scale height.

Explicitly, we will consider the following cases,
%
%
%

\begin{itemize}
\setlength{\itemindent}{-2em}
\setlength{\itemsep}{1.5ex}
  \setlength{\parskip}{0pt}
  \setlength{\parsep}{0pt}   
  
\item[] \textbf{PBO Case:} ``Powered Break-out'', SNe remnant superbubble reaches the gas disk scale height, $R_b = H$, \emph{before} the central star cluster ceases producing SNe, $t_{\rm BO} < t_{\rm SNe}$.

\item[] \textbf{CBO Case:} ``Coasting (unpowered) Break-out'', remnant reaches the gas disk scale height, $R_b = H$, \emph{after} the central star cluster ceases producing SNe, $t_{\rm BO} > t_{\rm SNe}$.

\item[] \textbf{CF Case:} ``Coasting (unpowered) Fragmentation'', the remnant fragments in the turbulent ISM (\ie the velocity of the shock-front falls below the turbulent velocity of the ISM), $v_b < \sigma$, \emph{before} reaching the gas disk scale height, $R_b(v_b = \sigma) < H$, \emph{after} the central star cluster ceases producing SNe, $t(v_b = \sigma) > t_{\rm SNe}$.

\item[] \textbf{PS Case:} ``Powered Stall'', bubble expansion stalls in the turbulent ISM, $v_b < \sigma$, \emph{before} reaching the gas-disk scale-height, $R_b(v_b = \sigma) < H$, \emph{before} the central star cluster ceases producing SNe, $t(v_b = \sigma) < t_{\rm SNe}$.
\end{itemize}

Table~\ref{table:cases} summarizes these outcome cases as a reference for the reader.
\subsection{PBO/PS Case: Successful Break-out or Fragmentation Before The End of Feedback ($t < t_{\rm SNe}$)}\label{sec:case0III}

When the superbubble remnant either successfully breaks out of the disk (\textbf{PBO case}), or stalls and fragments inside of it (\textbf{PS case}), \emph{before} the last SN explodes ($t < t_{\rm SNe}$), either $R_b = H$ \emph{and} $v_b > \sigma$ (\textbf{PBO case}), or $R_b < H$ \emph{and} $v_b = \sigma$ (\textbf{PS case}) occur when considering Eqs.~\ref{eq:rb} \& \ref{eq:vb}  $t < t_{\rm SNe}$.  If we assume that the gas disk that the remnant expands into is marginally Toomre-stable, \ie $\tilde Q_{\rm gas}=1$ (Eq.~\ref{eq:Q}), and solve Eqs.~\ref{eq:rb} \& \ref{eq:vb} to remove the time of break-out/fragmentation, we find the dividing condition between the cases to be
\begin{multline} \label{eq:case0III}
\tilde f_g = \frac{\sqrt{2}\pi G}{3} \frac{\Sigma_{\rm crit}}{(P/m_\star)_0} \left(\frac{4 \Omega t_{\rm SNe}}{2-\alpha}\right)^{1-\alpha} \frac{1}{\Omega} \; , \; \\ {\rm \textbf{(PBO/PS Case Border)}}
\end{multline}
with local gas fractions greater than this value belonging to the \textbf{PBO case}, where the remnant successfully breaks out of the disk (and drives \emph{powered} outflows), and local fractions less than it corresponding to the \textbf{PS case}, where the remnant fragments inside of the ISM while SNe are still occurring.

Inputting our fiducial values for various parameters (assuming $\alpha =0.46$) into Eq.~\ref{eq:case0III}, we see that the boundary between fragmentation and break-out is surprisingly close to Solar Circle conditions:
\begin{multline}\label{eq:case0III_nums}
\tilde f_g = 0.35 \left(\frac{\Sigma_{\rm crit}}{2800 \, {\rm M_\odot pc^{-2} }}\right) \left(\frac{3000 \, {\rm km/s}}{(P/m_\star)_0}\right) \\ \times \left(\frac{t_{\rm SNe}}{40 \, \rm Myr}\right)^{0.54}  \left(\frac{30 \, {\rm Gyr^{-1}} }{\Omega}\right)^{0.46} \; , 
\end{multline}
with the gas fraction along the boundary falling from $\sim$0.6 at $\Omega = 10$ Gyr$^{-1}$ to $\sim$0.2 at $\Omega = 100$ Gyr$^{-1}$.

Comparing similarly to the SFR-threshold for outflows study of \citet{Heckman2015}, we can estimate the star formation rate surface density (assuming we produce a star cluster according to this model once per local dynamical time), \ie $\dot\Sigma_\star \approx M_{\rm cl}\Omega/\pi H^2$. And using our $\tilde Q_{\rm gas} \approx 1$ assumption, derive, $\dot\Sigma_\star \approx 2 \sigma^2 \Omega^3 \tilde f_g/\pi^2 G^2 \Sigma_{\rm crit}$.  Inputting parameters typical of galactic centers, \eg $\sigma \approx 30$ km/s and $\Omega = 10^2$ Gyr$^{-1}$, and a $\tilde f_g \approx 0.2$ from Eq.~\ref{eq:case0III} for this $\Omega$, we find that the SFR at the \textbf{PBO/PS case} boundary for galactic center conditions is roughly
\be
\dot\Sigma_\star \approx 0.67 \frac{{\rm M_\odot}}{{\rm kpc^{2} \, yr }}\left(\frac{\tilde f_g}{0.2}\right) \left(\frac{\sigma}{30 \, {\rm km/s}}\right)^2   \left(\frac{\Omega}{10^2 \, {\rm Gyr^{-1}} }\right)^3 \; ,
\ee
which is agrees well with the SFR surface density threshold observed \citep{Heckman2015}.

A further requirement for the \textbf{PBO case} is that $t_{\rm BO} \leq t_{\rm SNe}$.  Solving Eq.~\ref{eq:rb} for $R_b = H$ \emph{and} $t \leq t_{\rm SNe}$, we find that for \emph{breakout} (not necessarily failure/fragmentation) to occur before $t_{\rm SNe}$,
\begin{multline} \label{eq:tbotsn}
\tilde f_g \geq \frac{\sqrt{2}\pi G}{3} \frac{\Sigma_{\rm crit}}{(P/m_\star)_0} \left(\frac{2-\alpha}{4 \Omega t_{\rm SNe}}\right) \frac{1}{\Omega} \; , \; \\ {\rm \textbf{(PBO/CBO Case Border)}}
\end{multline}
must hold.  That this requirement for local gas fractions is steeper than in Eq.~\ref{eq:case0III}, opens up a (small) wedge in $f_g$--$\Omega$ space for \textbf{CBO/CF cases}, where the superbubble remnants neither stall in the disk nor are able to break-out \emph{before} $t_{\rm SNe}$.  The \rev{point} that these lines (and the subsequent \textbf{CBO/CF case} and \textbf{CF/PS case} boundaries) intersect at is when $4\Omega t_{\rm SNe} = (2-\alpha)$, \emph{i.e.}, when the local dynamical time is $4/(2-\alpha)$ times the SNe duration timescale.

The remaining boundary for the \textbf{PS case} is found assuming that neither breakout nor stalling occurs \emph{before} $t_{\rm SNe}$, \emph{i.e.}, $R_b(t_{\rm SNe}) < H$ \emph{and} $v_b(t_{\rm SNe}) > \sigma$, and using the previous assumption that $\tilde Q_{\rm gas}=1$.  Solving the appropriate combination of Eqs.~\ref{eq:rb} \& \ref{eq:vb} yields,
\begin{multline} \label{eq:case23}
\tilde f_g \leq \frac{\sqrt{2}\pi G}{3} \frac{\Sigma_{\rm crit}}{(P/m_\star)_0} \left(\frac{4 \Omega t_{\rm SNe}}{2-\alpha}\right)^3 \frac{1}{\Omega} \; . \; \\ {\rm \textbf{(CF/PS Case Border)}}
\end{multline}
This steeply rising requirement for gas fraction demarcates the boundary between powered stall and coasting fragmentation (\textbf{CF/PS cases}) coming to the \rev{point} at $4\Omega t_{\rm SNe} = (2-\alpha)$ where all four cases meet.

By invoking marginal stability ($\tilde Q_{\rm gas}=1$), we find that the only `free' parameters in a galaxy (\ie those not tied to the IMF, stellar lifetimes, individual SN momentum yields, or cluster formation efficiency) in determining the outcome of the remnant evolution are the local gas fraction $\tilde f_g$ and (inverse) dynamical time $\Omega$.  The fact that stronger feedback, \pms, lowers the threshold for break-out is not particularly surprising. However, the whole directionality of the condition: that higher $\tilde f_g$ are required for break-out, and not lower gas fractions, \rev{\emph{may}} be counter-intuitive.  

\rev{The failure in intuition would lie in assuming that higher gas fractions \emph{only mean that clusters are more deeply embedded in the gas disk} (that there is a larger barrier to break-out).  But, in this feedback-regulated framework, the cumulative amount of feedback is connected to the mass of the central star cluster, and regions of galaxy disks with higher gas fractions produce larger Toomre-patches/GMCs.  These larger GMCs in turn produce more massive star clusters (\ie the Kennicutt-Schmidt relation is super-linear with total gas surface density): and so higher local gas fractions result in \emph{relatively} higher (per gas mass) integrated amounts of feedback injected into the gas disk from each feedback/star formation event.} 

\subsubsection{Flat SN Delay Time Distributions and Break-out}
In a universe where the IMF and stellar lifetimes have conspired to produce a perfectly flat distribution of SNe in time, \ie $\alpha = 0$, then Eq.~\ref{eq:case0III} reduces to
\be
\tilde f_{g, \alpha=0} = \frac{2\sqrt{2}}{3} \frac{\pi G \Sigma_{\rm crit} t_{\rm SNe}}{ (P/m_\star)_0} \approx 0.5 \; ,
\ee
where we have used the fiducial values for all of the feedback and star cluster formation efficiency parameters.  This quickly recovers some of our intuition regarding feedback, namely that spreading feedback across a longer duration $t_{\rm SNe}$ lessens its efficacy (here, increasing the required gas fraction for break-out).  Fascinatingly, however, the condition for break-out/stall becomes independent of $\Omega$ (\ie local dynamical time/galactic shear; presuming again, that $\tilde Q_{\rm gas}=1$ remains valid and we have a roughly uniform ISM with a definable gas scale height), and the ratio of the gas fraction to parameters regarding star cluster formation, feedback, and stellar lifetimes (all quantities related to much smaller scale phenomena) alone determines the outcome of clustered feedback events across galaxies.  In a $\alpha=0$ universe, however, the requirement of Eq.~\ref{eq:tbotsn} still holds, and continues to act as a boundary condition on the total amount of feedback momentum yielded per star cluster.  We explore the implications of a uniform delay time distribution in our model in Appendix~\ref{appendix:alpha0}.

\subsection{CBO/CF Case: Successful Break-out or Fragmentation After The End of Feedback ($t > t_{\rm SNe}$)}\label{sec:case12}
The cases where the superbubble either breaks out (\ie coasts out) of the disk or fragments inside the disk \emph{after} the last SN explodes ($t > t_{\rm SNe}$) share two sides of the same inequality.  Taking the conditionals in Eqs.~\ref{eq:rb} \& \ref{eq:vb} where $t > t_{\rm SNe}$, and either demanding that: $R_b = H$ \emph{and} $v_b > \sigma$ (\textbf{CBO case}), or $R_b < H$ \emph{and} $v_b = \sigma$ (\textbf{CF case}), yields a condition dividing the two cases of
\begin{multline} \label{eq:case12}
\tilde f_g = \frac{\sqrt{2}\pi G}{3} \frac{\Sigma_{\rm crit}}{(P/m_\star)_0} \frac{1}{\Omega} \; , \; \\ {\rm \textbf{(CBO/CF Case Border)}}
\end{multline}
with local gas fractions $\tilde f_g$ greater than this value belonging to the \textbf{CBO case}, where the remnant successfully coasts out of the disk, and $\tilde f_g$ less than it corresponding to the \textbf{CF case}, where the coasting remnant fragments inside of the gas disk.
Again, like the division between the previous \textbf{PBO/PS cases}, this has invoked the condition that the disk be marginally stable $\tilde Q_{\rm gas}=1$ as the condition for the disk that the remnant expands into.  Interestingly, here the division between coasting (unpowered) break-out or fragmentation does not relate at all to the duration of feedback $t_{\rm SNe}$, nor to the distribution of it in time, $\alpha$: the only condition that divides these cases is the ratio between local gas fraction $\tilde f_g$ and dynamical time $\Omega$, as the cumulative feedback \emph{and} disk conditions are connected between assumptions of $\tilde Q_{\rm gas}=1$ as well as the adopted star cluster formation efficiency/feedback strength model.

Figure~\ref{fig:case} shows how the various outcomes relate to each other in $\tilde f_g$--$\Omega$ space.  Owing to the fact that \textbf{PBO/PS cases} occur \emph{before} $t_{\rm SNe}$, these cases supersede the coasting/unpowered outcomes of \textbf{CBO/CF cases}.  Evidently, there is only a \emph{very} small region in $\tilde f_g$--$\Omega$ space where break-out occurs \emph{after} the last SNe occurs, at least with the fiducial values assumed in our model (see Appendix~\ref{appendix:parameters} for an exploration of the meaningful input parameter space of the model; for any reasonable variation of parameters, including $\alpha$, the \textbf{CBO case} remains rare).  Thus, regions that fall below either the lines of Eq.~\ref{eq:case0III} \emph{or} \ref{eq:case12} remain in the classical model of SNe feedback in disks, where the feedback from SNe is entirely retained by the ISM (\textbf{CF/PS cases}), and regions above the lines of Eq.~\ref{eq:case0III} \emph{and} \ref{eq:tbotsn} exist in a space where feedback is partially deposited locally in disks, but some fraction of the feedback (that which occurs between $t_{\rm BO}$ and $t_{\rm SNe}$) drives outflows/fountains (\textbf{PBO case}).  Resultantly, this model implies that \emph{almost all} outflows/fountains are \emph{powered}, and thus have a high energy-loading phase. We discuss the potential effects on the case/outcome boundaries of a clumpy/inhomogeneous ISM in \S~\ref{sec:conservative}. 

\subsection{\rev{Comment on the Extreme Rarity of CBO \& CF Cases (Coasting Outcomes) in Disk Galaxies}}\label{sec:critpt}

In deriving the boundaries between cases, we found that there exists a \rev{``}critical\rev{"} point where all four of the cases intersect at 
\be \label{eq:fcrit}
\tilde f_g^{\rm crit} = \frac{\sqrt{2}\pi G \Sigma_{\rm crit}}{3 (P/m_\star)_0 \Omega^{\rm crit}} \; \;  \& \; \; \Omega^{\rm crit} = \frac{2-\alpha}{4 t_{\rm SNe}}  \; .
\ee
This gas fraction resembles something of a `$Q$-critical', in the form of $Q_{\rm crit} \approx \sqrt{2}(P/m_\star)\Omega^{\rm crit}/\pi G \Sigma_{\rm crit}$, relating to the maximum stabilizable surface density for SNe feedback \citep[see][for their discussion of surface density where SNe \rev{begin} to fail to regulate star formation]{Grudic2019}.  \rev{We can frame this $Q_{\rm crit}$ as the Toomre-Q value associated with the limit of supportive SN feedback, where the turbulent/momentum support term is the `natural' velocity/pressure term associated with SN feedback $(P/m_\star)$.  $\Sigma_{\rm crit}$ is then the surface density above which $Q_{\rm crit} < 1$, implying the failure of SN feedback as a mechanism to resist fragmentation and collapse.}   And so, this \rev{``}critical\rev{"} point is thus at $\tilde f_g = 2/3Q_{\rm crit}$.

However, for disk galaxies at late times, the far more difficult condition is to find regions of the disk with such long dynamical times to fall into \textbf{CBO} or \textbf{CF cases} at all.  For $t_{\rm SNe} = 40$ Myr, and a MW-like rotation curve of 220 km/s, the radius at which this condition is met is $R = 4 v_c t_{\rm SNe}/ (2-\alpha) \approx 23.4$~kpc, which is beyond the edge of the (high-mass) star-forming disk in our galaxy \citep{Djordjevic2019}.  In a disky dwarf galaxy like M33, this radius is considerably closer at $R\approx 13.5$~kpc \citep[taking M33's $v_c = 127$ km/s;][]{Corbelli2000}, but still perhaps beyond the edge of the star-forming disk ($\sim$7~kpc, \citealt{Verley2009}).  Consequently, with such long dynamical times not appearing to correspond to regions of active star formation in disk galaxies at late times \citep{Casasola2017}, we would \emph{extremely rarely} expect to see the \textbf{CBO} or \textbf{CF case} outcomes for superbubbles: fragmentation or break-out will \emph{almost always} be powered ($t_{\rm BO}<t_{\rm SNe}$).  That is, beyond the fact that the region of parameter space corresponding to the \textbf{CBO case} is incredibly narrow, few star-forming disks are likely to inhabit that part of $\tilde f_g$--$\Omega$ space in the first place.

\section{Implications for Turbulence Driving Scale: CF \& PS Cases}\label{sec:turbscale}
\begin{figure}
	\includegraphics[width=0.97\columnwidth]{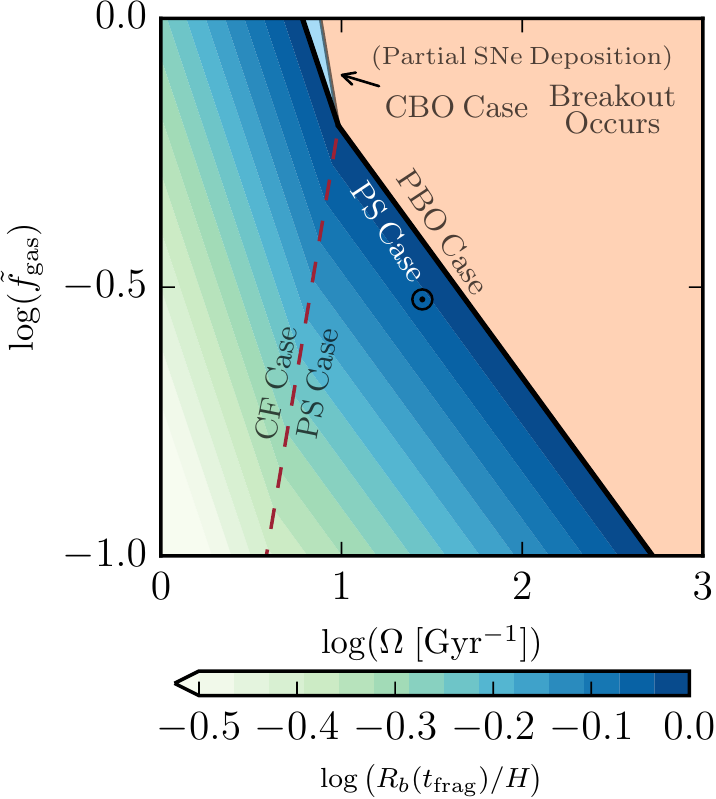}
    \caption{Fragmentation radius as a fraction of disk scale height of superbubbles that fail to break out of the disk (\textbf{CF/PS cases}), in gas fraction--dynamical time phase space, following Eqs.~\ref{eq:fragscale-case2} \& \ref{eq:fragscale-case3}, with 
    case boundary lines as in Fig.~\ref{fig:case}.  Solid orange and light blue regions denotes cases where remnant successfully breaks out of the disk (\textbf{PBO/CBO cases}), and with a solid black line separating those regions from the area of parameter space corresponding to bubble fragmentation.  Dashed cardinal line indicates division between \textbf{CF} (coasting fragmentation) and \textbf{PS} (powered stall) \textbf{cases}. For coasting/unpowered fragmentation (\textbf{CF case}), the shock front velocity comes into pressure equilibrium and fragments when $v_b = \sigma$, and no further SNe momentum injection occurs.  In the \textbf{PS case}, however, when $v_b = \sigma$ occurs \emph{before} $t_{\rm SNe}$, then the shock fragments at that radius while SNe momentum continues to be injected at that scale for a time $t_{\rm SNe} - t_{\rm frag}$. }
    \label{fig:fragscale}
\end{figure}
In the event that the superbubble stalls and fragments inside the disk, this model predicts that the shock front will do so when $v_b = \sigma$ before $R_b$ reaches $H$.  For example, in Fig.~\ref{fig:evolve} the $\tilde f_g = 0.1$ model superbubble stalls near $R_b/H \sim 0.6$, well before reaching the edge of the disk. We then might expect that the driving scale of turbulence in the ISM corresponds to this stall/fragmentation scale.  The radius at which this occurs at depends critically on whether or not this occurs before/after $t_{\rm SNe}$.

For the \textbf{PS case}, where $t<t_{\rm SNe}$, we solve for this radius by solving for the stall/fragmentation time $t_{\rm frag}$ where $v_b(t_{\rm frag}) = \sigma$ in Eq.~\ref{eq:vb}, and then solving for the stall/fragmentation radius $R_b(t_{\rm frag})$ using Eq.~\ref{eq:rb}, and assuming $\tilde Q_{\rm gas}=1$ to find,
\begin{multline} \label{eq:fragscale-case3}
R_b(t_{\rm frag}) = H \left[ \frac{(P/m_\star)_0}{\Sigma_{\rm crit}} \frac{3 \Omega \tilde f_g}{\sqrt{2}\pi G} \right]^\frac{1}{2+\alpha} \left(\frac{2-\alpha}{4\Omega t_{\rm SNe}} \right)^\frac{1-\alpha}{2+\alpha} \;  \\ \textbf{(PS Case)}.
\end{multline}
Here we can see that this predicts that break-out occurs along the same line in $\tilde f_g$--$\Omega$ space as described by Eq.~\ref{eq:case0III}, having the same dependence on local gas fractions and dynamical times.

The matter is simpler for \textbf{CF case}, where $t>t_{\rm SNe}$ and thus all of the SNe momentum has been injected regardless of its particular distribution in time (no $\alpha$ dependence).  Solving Eqs.~\ref{eq:rb} \& \ref{eq:vb} again, with $\tilde Q_{\rm gas}=1$, for the time after $t_{\rm SNe}$ where $v_b(t_{\rm frag}) = \sigma$ and $R_b(t_{\rm frag}) < H$, we find,
\begin{multline} \label{eq:fragscale-case2}
R_b(t_{\rm frag}) = H \left[ \frac{(P/m_\star)_0}{\Sigma_{\rm crit}} \frac{3 \Omega \tilde f_g}{\sqrt{2}\pi G} \right]^\frac{1}{3} \;  \\ \textbf{(CF Case)}.
\end{multline}
Once again, we see that contours of constant $R_b(t_{\rm frag})$ in the \textbf{CF case} region have the same slope in $\tilde f_g$--$\Omega$ space as described by the \textbf{CBO/CF case} boundary (Eq.~\ref{eq:case12}).  Further, we see that there is no discontinuity along the \textbf{CF/PS case} boundary described by Eq.~\ref{eq:case23}.

Figure~\ref{fig:fragscale} shows the fragmentation radius, as a function of disk scale height, across the parameter space of $\tilde f_g$--$\Omega$, highlighting how the fragmentation radius smoothly grows to scale height along both the powered and coasting break-out boundaries. Further, we thus predict that there is a driving scale, which is a fraction of disk scale height, for SNe-driven turbulence in disks that depends on a combination of the strength of stellar feedback/star cluster formation efficiency and local disk properties.   
Highly resolved velocity maps of dense gas in nearby disk galaxies may be able to detect this signature momentum injection scale.
\section{Implications for Kennicutt-Schmidt: PBO Case}\label{sec:KS}

\begin{figure}
	\includegraphics[width=0.97\columnwidth]{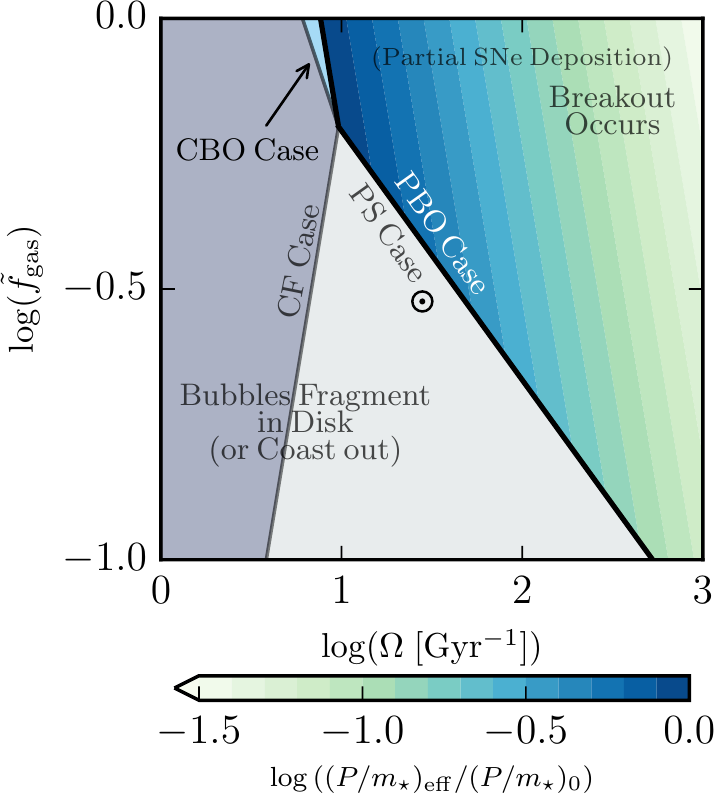}
    \caption{Ratio of `effective' to fiducial feedback strength of superbubbles that successfully break out of the disk \emph{before} $t_{\rm SNe}$ in gas fraction--dynamical time phase space, following Eq.~\ref{eq:pmseff}, with 
    case boundary lines as in Fig.~\ref{fig:case}.  Solid light blue, blue-gray, and light gray regions denotes cases where remnant either fragments within the disk (\textbf{CF/PS cases}) or coasts out of the disk (\textbf{CBO case}), with solid black line separating those regions from the outcomes of the \textbf{PBO case}.  As there is a smooth transition between the bubble coasting out of the disk in \textbf{CBO case} and \emph{just barely} breaking out with some SNe yet to occur, there is no discontinuity along the \textbf{PBO/CBO case} boundary (\emph{i.e.,} $(P/m_\star)_{\rm eff}/(P/m_\star)_0=1$ there). However, in the case of fragmentation \emph{before} $t_{\rm SNe}$ (\textbf{PS case}), a significant fraction of the SNe momentum injection can occur \emph{after} fragmentation.  And so, there is a discontinuity along the \textbf{PBO/PS case} boundary: $(P/m_\star)_{\rm eff} \neq (P/m_\star)_0$ there (once breakout occurs the superbubble may have `a lot left in the tank' by way of SNe to occur).  }
    \label{fig:effPMS}
\end{figure}

In the case where a superbubble breaks out of the galaxy disk \emph{before} all the SNe from a given star cluster occur (\textbf{PBO case}), the momentum from those remaining SNe are not directly deposited in the ISM: some fraction of the `leftover' feedback momentum is instead directed into either galactic outflows or fountains.  If we are to assume that the turbulent ISM on large scales in a galaxy is regulated by the feedback momentum following star formation, then the yield of feedback into the ISM per mass of stars formed is crucial to any associated model.  `Classical' models of feedback in disks assume that the momentum yield is simply the $t > t_{\rm SNe}$ case of Eq.~\ref{eq:psne}, \ie that the normalized `strength of feedback' is always $(P/m_\star)_0$.  However, in the case of a successful powered superbubble break-out, then the `effective' strength of feedback $(P/m_\star)_{\rm eff}$ is \emph{less} if we consider all of the momentum yielded from SNe that occur \emph{after} break-out to be `lost' to outflows/fountains.  The yield of feedback into the ISM from this model is then,
\be \label{eq:pmseff_0}
(P/m_\star)_{\rm eff} = (P/m_\star)_{0} \left( \frac{t_{\rm BO}}{t_{\rm SNe}} \right)^{1-\alpha} \; .
\ee
Solving Eq.~\ref{eq:rb} for $R_b = H$ and $t_{\rm BO} = t < t_{\rm SNe}$, and using the simplifying assumption that $\tilde Q_{\rm gas}=1$, we find,
\be \label{eq:pmseff}
\frac{(P/m_\star)_{\rm eff}}{(P/m_\star)_{0}} = \left[ \frac{\sqrt{2}\pi G}{3} \frac{\Sigma_{\rm crit}}{(P/m_\star)_{0}} \frac{2-\alpha}{4 \Omega t_{\rm SNe}} \frac{1}{\tilde f_g\Omega} \right]^{\left(\frac{1-\alpha}{2-\alpha}\right)} \; .
\ee
Figure \ref{fig:effPMS} plots this quantity in gas fraction--dynamical time space, where valid (\ie for the \textbf{PBO case}).  Owing to the fact that the effective strength of feedback is only weakly dependent on local gas fraction, $(P/m_\star)_{\rm eff} \propto \tilde f_g^{-0.35}$, the strongest determinant of the effectiveness of feedback in driving turbulence across disks is the local dynamical time/relative amount of shear in the form of $(P/m_\star)_{\rm eff} \propto \Omega^{-0.7}$.  Hence, for regions with dynamical times on the order of the stellar evolutionary timescales, \ie $\Omega \approx 1/t_{\rm SNe}$, $(P/m_\star)_{\rm eff} \approx 0.42 (P/m_\star)_0$, prescribing a realistic lower limit on how weak feedback \emph{within} disks can become in this model. 

We can apply the derived effectiveness of feedback derived in Eq.~\ref{eq:pmseff} to models of feedback-regulated disks found in \citet{Faucher-Giguere2013}, \citet{Hayward2017} or \citet{Orr2018}, where rate of dissipation of momentum in supersonic turbulence (decaying on a disk crossing time, $t_{\rm diss} \sim H/\sigma \sim 1/\Omega$) is balanced by the injection of feedback momentum following star formation.  At its core, this framework for feedback-regulation takes the form $\dot\Sigma_\star (P/m_\star) \approx \sqrt{3}\sigma \Sigma_g \Omega/2$ (see \citealt{Orr2019}, Eq.~5).  This model is indistinguishable from one assuming that disks are in hydrostatic equilibrium, with the gravitational weight of the disk balanced by the momentum flux of feedback, so long as $\tilde Q_{\rm gas}=1$. Hence, substituting for $\sigma \Omega$, the rate of star formation derived in this model becomes: $\dot\Sigma_\star \approx \sqrt{3} \pi G \Sigma_g^2/2\sqrt{2}\tilde f_g (P/m_\star)$.  However, these models often assume that $(P/m_\star)$ is a constant, informed by simulations of individual SN explosion simulations where the terminal momentum of the shockwave is very weakly dependent on local ISM conditions \citep[\eg][]{Martizzi2015}.  By substituting in our derived \emph{effective} strength of feedback $(P/m_\star)_{\rm eff}$, we derive a \textbf{PBO case} KS relation of the form,
\begin{multline} \label{eq:KS}
\dot\Sigma_\star \approx \frac{\sqrt{3} \pi G}{2\sqrt{2}} \frac{\Sigma_g^2}{\tilde f_g (P/m_\star)_{0}} \\ \times \left[ \frac{\sqrt{2}\pi G}{3} \frac{\Sigma_{\rm crit}}{(P/m_\star)_{0}} \frac{2-\alpha}{4 \Omega t_{\rm SNe}} \frac{1}{\tilde f_g\Omega} \right]^{-\left(\frac{1-\alpha}{2-\alpha}\right)} \; .
\end{multline}
The major implication being that as dynamical times shorten towards galactic centers, and outflows/fountains become more commonplace, the effective strength of feedback falls causing the appearance of a steeper Kennicutt-Schmidt relation.  
This being separate from the steepening from a gas-poor slope (\ie $\dot\Sigma_\star \propto \Sigma_g$ when $\tilde f_g \rightarrow 0$) to a gas-rich slope (\ie $\dot\Sigma_\star \propto \Sigma_g^2$), that may occur in galactic outskirts.  Indeed this may even connect to Kennicutt-Schmidt observations of starburst galaxies at higher redshift, where high gas fractions and short dynamical times may conspire towards reducing the effectiveness of feedback within galaxies.  \rev{The notion that high-redshift galaxies are \emph{anomalously efficient} given their relatively small inferred cold molecular gas reservoirs (from low $X_{\rm CO}$ values; \citealt{Genzel2011, Bolatto2013}) may be incorrect: feedback could simply be a less effective regulator given the local ISM conditions.}

\rev{As one considers the changing environments of star formation in galaxies, from their outskirts to their centers:} the transitions from gas rich (and perhaps predominantly thermally supported) outskirts to turbulently supported molecular ISM disks \rev{(which may fully contain their superbubbles)} to central \rev{nuclear} regions with \emph{powered} superbubble breakouts may correspond to the \rev{changes in power-law slope} seen in \rev{the spatially resolved} Kennicutt-Schmidt \rev{relation} \citep{Bigiel2008, Hayward2017}.
\section{Simple Outflow Model: PBO \& CBO Case }\label{sec:outflows}

\begin{figure}
	\includegraphics[width=0.97\columnwidth]{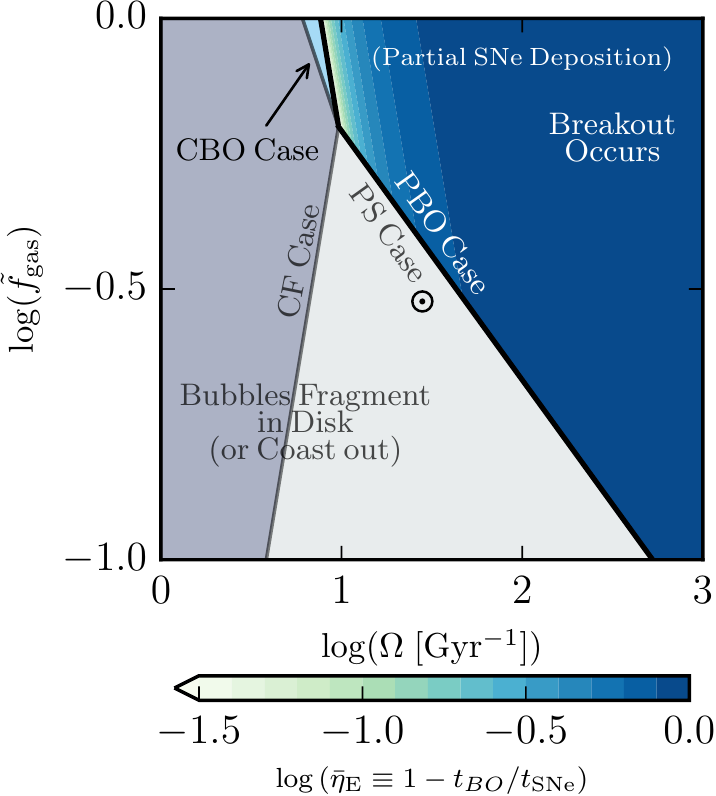}
    \caption{Open channel fractional duration (Eq.~\ref{eq:topenchannel}, $1 - t_{\rm BO}/t_{\rm SNe}$, analogously the maximal average energy loading) of outflows from \textbf{PBO case} superbubble outcomes in gas fraction--dynamical time phase space, following Eqs.~\ref{eq:case0III} \& \ref{eq:case12}, with 
    case boundary lines as in Fig.~\ref{fig:case}.  The boundary between \textbf{PBO} \& \textbf{CBO cases} is a smooth transition between the remnant having no open-channel time, and there being a period where a channel is open to the near-CGM for the subsequent SNe to expand into.  The boundary between \textbf{PBO} \& \textbf{PS cases} represents a discontinuity between having no open channel (SNe fragment in the disk) and having significant open channel durations, with the duration rapidly approaching $t_{\rm SNe}$. Alternately, the open-channel fractional duration can be viewed as the maximal average energy loading factor for outflows (the average energy loading factor would be reduced by any ISM entrained during the open-channel phase).}
    \label{fig:effetaE}
\end{figure}

\begin{figure}
	\includegraphics[width=0.97\columnwidth]{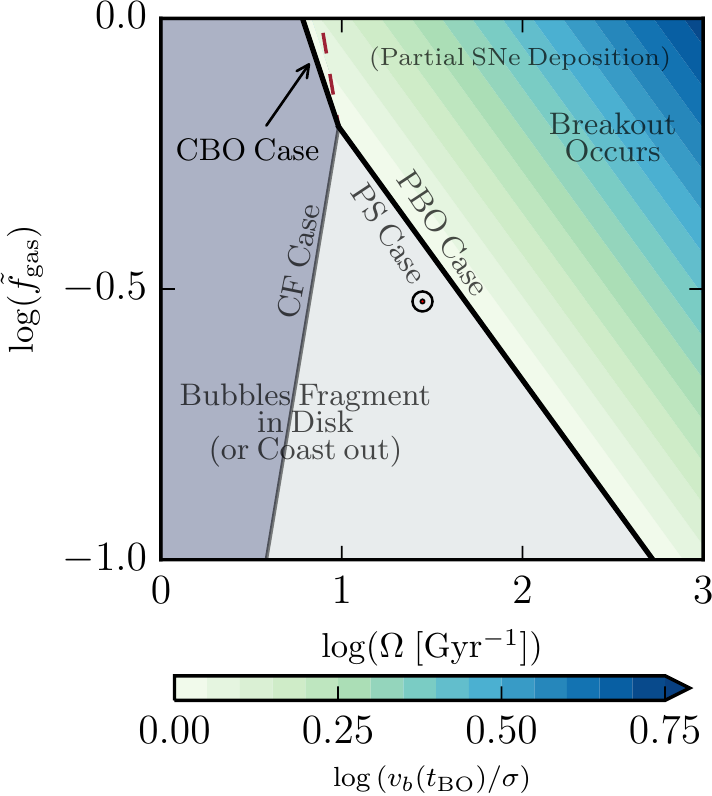}
    \caption{Superbubble velocity at break-out as a fraction of velocity dispersion for superbubbles that break out of the disk (\textbf{PBO/CBO cases}), in gas fraction--dynamical time phase space, following Eqs.~\ref{eq:vbo}, with 
    case boundary lines as in Fig.~\ref{fig:case}.  Blue-grey and light\rev{-}grey regions denotes cases where remnant fails to breaks out of the disk (\textbf{CF/PS cases}), and with a solid black line separating those regions from the area of parameter space corresponding to break-out.  Dashed cardinal line indicates division between \textbf{PBO} (powered break-out) and \textbf{CBO} (coasting break-out) \textbf{cases}. For coasting/unpowered break-out (\textbf{CBO case}), the shock front barely coasts out of the disk $v_b(t_{\rm BO}) \approx \sigma$, and no `hot' wind follows.  In the \textbf{PBO case}, however, the shock front can leave with $v_b(t_{\rm BO})$ several times $\sigma$, with a `hot wind' following for a time $t_{\rm SNe} - t_{\rm BO}$.}
    \label{fig:vbo}
\end{figure}

In this model, the threshold for outflows in galaxies is not strictly in terms of a gas or star formation rate surface density (\eg \citealt{Heckman2015}, though we show in \S~\ref{sec:case0III} that our predicted SFRs are in rough agreement with observed SFR `thresholds' for outflows), instead it is tied to the local galactic disk properties, namely, the local gas fraction $\tilde f_g$ and orbital dynamical time $\Omega$.  The complexity of SN-driven outflows themselves, with their attendant questions surrounding cloud crushing, entrainment, etc.~\citep{Scannapieco2015, Schneider2017, Zhang2017, Gronke2020, Abruzzo2021}, is well-beyond the scope of the simple model we present here.  However, we can speak in phenomenological terms regarding outflows predicted by this model\rev{: specifically the hot energy-carrying phase only. We reference \citet{Fielding2022} for an exploration in detail of the neglected effects, and a realistic model for multiphase galactic winds.}

A key parameter is the open-channel time \rev{while SNe are occuring}, this being the difference between the time of break-out $t_{\rm BO}$ and the time of the last SN $t_{\rm SNe}$, assuming that the \rev{duration of the outflow driving ends at the time of} the last supernova. \rev{This is a simplification as certainly the ejecta of the last supernovae take additional time to propagate. Additionally, the channel closing time, being at least a sound crossing time, is expected to be longer than $t_{\rm SNe}$.}  \rev{Regardless, }having calculated $t_{\rm BO}/t_{\rm SNe}$ in \S~\ref{sec:KS}, we can construct the fractional open channel time for the \textbf{PBO case}, $1 - t_{\rm BO}/t_{\rm SNe}$,
\be \label{eq:topenchannel}
\overline{\eta_{\rm E}} \equiv 1 - \frac{t_{\rm BO}}{t_{\rm SNe}} =  1- \left[ \frac{\sqrt{2}\pi G}{3} \frac{\Sigma_{\rm crit}}{(P/m_\star)_{0}} \frac{2-\alpha}{4 \Omega t_{\rm SNe}} \frac{1}{\tilde f_g\Omega} \right]^{\frac{1}{2-\alpha}} \; .
\ee
This fractional open channel time sets an approximate upper limit on the \rev{energy loading factor averaged over the lifetime of the star cluster} $\overline{\eta_{\rm E}}$, which is defined to be the ratio of the energy that escapes the ISM in the form of an outflow to the total energy that has been injected by the SNe. In reality, the energy loading factor will not be quite as high because additional energy will be lost as ISM turbulence pushes fresh cold material into the path of the outflow. This continued injection of material into the outflow channel will provide most of the mass loading of the resulting outflow. Much of the cold material will be fully shredded and added to the hot phase, while some of the cold material may survive in the form of clouds that are entrained by the hot flow \citep[\eg][]{Gronke2020, Fielding2022}. This estimate for $\overline{\eta_{\rm E}}$, therefore, sets the expected strength of the outflow both in terms of its ability to eject material from the ISM and to heat the surrounding CGM and prevent future accretion \citep{Li2020}.

The general picture presented here of a superbubble shell sweeping up the ambient ISM, and then breaking out of the gas disk, has the immediate prediction that we would expect there to first be a dense `cold cap' that is expelled, with roughly the metallicity/enrichment of the gas disk\footnote{Except for very-low-metallicity gas, the metal mass in the ambient gas reservoir should dwarf the mass of metals returned to the ISM by even the largest star cluster \eg presuming a $\sim$$10^6$~\msolt GMC yields a $\sim$$10^4$~\msolt cluster, the model of \citet{Agertz2013} argues the cluster will return only $\sim$$10^2$ \msolt of metals (they assume nearly identical IMF slopes and SNe per mass of stars formed as we do here), whereas that GMC has a mass of $\sim$$10^6 Z_{\rm GMC}$ \msolt in metals.  Thus an individual star cluster will hardly affect the overall metallicity of the ISM for gas abundances $\gtrsim 10^4 Z_\odot$, which is readily achieved in star-forming galaxies before at least $z \approx 3$ \citep{Zahid2013, Izotov2015}.}.  Once the `cap' is lifted off as the remnant breaks out of the disk, this model then predicts that an open channel (lasting for $t = t_{\rm SNe} - t_{\rm BO}$) provides a means for hot ejecta from subsequent SNe (in the \textbf{PBO case}) to directly launch out of the disk with a metallicity equal to that of the SNe themselves.  \textbf{CBO case} superbubbles, having barely lifted the dense gas to the edge of the gas disk, would simply adiabatically deflate their hot gas into the near-CGM, rather than actively powering hot winds.  Again, this is ignoring the many complexities of metal mixing and dense gas entrainment doubtlessly involved. Though neither explicitly followed gas metallicity/enrichment from SNe, the superbubble simulations of \citet{Kim2016} and tall-box simulations of \citet{Kim2017} exhibited these lifted dense (and reasonably cold) `caps' followed by hot gas escaping.  We can calculate the velocity of the superbubble at break-out by inserting the appropriate $t_{\rm BO}$ in Eq.~\ref{eq:vb} (we calculate the $t_{\rm BO}$ of the \textbf{CBO case} by solving the $t > t_{\rm SNe}$ case of Eq.~\ref{eq:rb}), and we find that the velocity of the accelerated `cold cap' is
\be \label{eq:vbo}
\frac{v_b(t_{\rm BO})}{\sigma} =
\begin{cases}
\!\begin{aligned}
       &\frac{2 - \alpha}{4 \Omega t_{\rm SNe}} \left[ \frac{3}{\sqrt{2} \pi G} \frac{(P/m_\star)_0}{\Sigma_{\rm crit}} \tilde f_{\rm gas} \frac{4 \Omega t_{\rm SNe}}{2 - \alpha} \right]^{\frac{1}{2 - \alpha}}\\
       &  \qquad \qquad \qquad \qquad \qquad \qquad \textbf{PBO case} 
    \end{aligned} & \\

\!\begin{aligned}
       &\frac{3}{\sqrt{2} \pi G} \frac{(P/m_\star)_0}{\Sigma_{\rm crit}} \tilde f_{\rm gas} \Omega\\
       &  \qquad \qquad \qquad \qquad \qquad \qquad \textbf{CBO case} 
    \end{aligned} & \\
\end{cases}
\ee
As with scalings surrounding the \textbf{CF case}, we see that the velocity of the `cold cap' in the \textbf{CBO case} has no dependence on $\alpha$ or $t_{\rm SNe}$, since only the integrated amount of feedback matters here.  Figure~\ref{fig:vbo} shows Eq.~\ref{eq:vbo} plotted in $\tilde f_{\rm gas}$--$\Omega$ space for \textbf{PBO} \& \textbf{CBO cases}. With values of $\log(v_b(t_{\rm BO})/\sigma)$ `near' the \textbf{PBO/PS case} boundary not exceeding $\sim$0.3, for typical Milky Way disk conditions, we would expect to only see cold fountains launching with few tens of km/s velocities, similar to those seen by \citet{Kim2016}.
 
The velocity at breakout is, however, not reflective of the expected velocity of either the hot phase or the colder phases of the resulting outflow. The post-breakout velocity of the hot volume filling phase of the outflow will depend on the degree to which additional ISM material is mixed in over the course of the open channel time (with higher mass loading leading to lower velocities and lower energy fluxes due to radiative losses). Furthermore, the velocity of the cold material driven out of the galaxy by such a feedback event will depend primarily on the details of cooling and mixing between the hot wind and the cold clouds \citep[\eg][]{Gronke2020, Abruzzo2021}. Recently, \citet{Fielding2022} presented a model for the interaction and evolution of a hot wind with embedded cold clouds and found that, consistent with observations, in many cases cold material that leaves the ISM at a few tens of km/s can be accelerated to hundreds or thousands of km/s within several kpc of the galaxy.

\section{Discussion}\label{sec:discussion}
\subsection{Effects of Early Feedback, Gas Turbulence, and the Vertical Distribution of Clusters} \label{sec:conservative}

Early feedback from OB stars in young clusters, in the form of stellar winds, photo-electric heating, and photo-ionizing radiation, is certainly critical in setting the local star formation efficiency \citep[\eg][]{Dale2014, Grudic2018, Li2019, Kim2020, Smith2020arx}, and that effect from early feedback is already implicitly accounted for here through the star cluster formation model (\S~\ref{sec:SCFM}).  However, another important implicit assumption also made in our model is that this early feedback removes the dense gas from the birth sites of the massive stars which will undergo core-collapse.  This is necessary for SNe to expand beyond their progenitor clouds, as SN explosions do not appear to dramatically affect very dense gas in GMCs \citep{Seifried2018, Lucas2020}, as it is generally strongly gravitationally self-bound and already collapsing.  In addition to affecting the density of the immediate sites of the SNe, this early feedback has also been shown to be key in carving out low-density channels that SNe follow \citep[][Appel et al. \emph{in prep.}]{Rogers2013, Walch2015, Lucas2020}.   

Further in this vein, supersonic gas turbulence alone may aid superbubble remnants in disk break-out.  Turbulence naturally forms low density channels that the expanding remnant may tend to follow vertically out of the disk \citep{Korolev2015}.  This was seen in the individual-SN simulations of \citet{Martizzi2015}, in which there was a roughly factor of two increase in the size of SN remnants in an inhomogeneous versus homogenous ISM, as SNe expanded along (low-density) channels of least resistance. \rev{The clustered SN simulations of \citet{Fielding2018} also corroborate this point, with them finding that a superbubble in a turbulent inhomogeneous medium grew to be about 50\% larger.} And in the event that an expanding bubble does encounter a patch of dense bound gas, the most likely outcome is that the bubble will simply sweep around it \citep{Seifried2018}. 

Considering again the difficulty for SN to drive turbulence in (or generally affect) very dense gas, we might worry that the assumption that the bubble sweeps out an order unity fraction of the ISM is very wrong \ie that we should replace $\bar \rho_g$ in Eq.~\ref{eq:pbub} with some `effective' density that accounts for the fraction of the ISM the SNe are able to sweep out.  Helpfully, the model of \citet{Burkhart2018a} and \citet{Burkhart2019} considers the fraction of gas in a supersonically turbulent ISM, otherwise following a log-normal density PDF, that is gravitationally self-bound and in the (actively collapsing) power-law tail of the gas density PDF, $f_{\rm dense}$. From this, we can estimate the non-bound fraction of gas that SNe readily affect, $1-f_{\rm dense}$.  Assuming a power-law slope of 1.5 (consistent with the behavior of ideal collapsing isothermal cores), a sound speed $\sim$0.3 km/s in the cold dense gas, and a turbulence driving parameter $b\approx 0.5$ ($b \approx 0.3$ being purely solenoidally driven and $b \approx 1$ purely compressively driven), their model predicts $f_{\rm dense} \approx 0.06$ at $\sigma = 5$ km/s falling to $f_{\rm dense} \approx 0.015$ at $\sigma = 15$ km/s (cf. the behavior of $f_{\rm dense}$ in Figure~5 of \citealt{Burkhart2019}, noting the different parameters used there, which suggest that bound gas fractions may be higher-- on the order of tens of percent).  Relating directly to the fact that star formation is inefficient, self-bound gas is at most few percent of the mass of the turbulent ISM, and our assumption that an expanding superbubble affects \emph{most} of the ISM appears valid.  

Lastly here, the assumption that star clusters form at exactly the disk midplane does not necessarily always hold either.  Scatter in the initial vertical distribution of young star clusters would result in there being less material for a superbubble to sweep up before breaking out of the `near' side of the disk, thus making breakout somewhat easier. That said, except for star clusters that form at `extreme' vertical heights relative to the galaxy midplane (\ie $z_{\rm cluster} \sim H$), for which the full-thickness of the gas disk may represent an insurmountable obstacle for the blastwave on one side (with a negligible gas column on the other side), heavily asymmetric outflows above and below the disk are unlikely given the excess momentum for successful \textbf{PBO case} superbubbles, though this is decreasingly the case for regions approaching the \textbf{PBO/CBO case} boundary (see \S~\ref{sec:KS})\footnote{Galactic outskirts, with their predicted-to-be-weak break-outs, accordingly may then be `interesting', in a sense, as the likeliest setting for asymmetric outflows (see \S~\ref{sec:galevolve}).}.

And so, to some varying extent, early feedback, supersonic turbulence, and the distribution of young clusters vertically ought to aid superbubble break-out.  Rather than moving the \textbf{PBO/PS case} and \textbf{PBO/CBO case} divisions to lower $\tilde f_g$, the boundaries might be smeared out in $\tilde f_g$--$\Omega$ space, constituting `soft' (and generally conservative) thresholds for break-out.

\subsection{\rev{Magnetic Fields and Breakout}} \label{sec:magfields}

\rev{Though we neglect the effects of magnetic fields in our minimal-physics model, recent works by \citet{Kim2018a} and \citet{Kim2020a} observe powerful outflows consistent with our findings here in simulations that do include magnetic fields.  In general, it is seen that magnetic fields do not strongly inhibit the expansion of bubbles or outflows, but rather shape them, as the field energy density is strongly subdominant to either the kinetic or thermal terms in these environments.  Magnetic fields, however, may have dramatic implications for the entrainment of cold material into winds and the lifetimes of cold clouds in hot outflows \citep{McCourt2015}.}

\subsection{IMF Sampling and Star Formation Triggers} \label{sec:modelfail}

\citet{Grudic2018} validated their cluster formation efficiency model to surface densities as low as $\sim $20 \mpcsqt.  They found that the modeled effects of prompt feedback processes, which affect the instantaneous SFR of GMCs and the cluster formation efficiency of individual cluster formation events, do not appear to dramatically change in their efficacy down to the typical gas surface density of MW clouds (for MW clouds, see \citealt{Heyer2015}).  In that regime, the IMF is well-enough sampled such that the average specific strength of photoionization/protostellar winds is roughly constant. 

In the sub-$10^3$ \msolt stellar mass cluster limit, we approach a poorly sampled IMF that poses problems to both the assumptions regarding the strength/efficacy of prompt feedback and that we can treat the time delay distribution of SNe as continuous.   To the extent that there are $\lesssim 10$ SNe occurring over the $t_{\rm SNe}$ period for a $10^3$ \msolt cluster, it is clear that modeling them as stochastic events becomes more appropriate.  We do not, however, gain clarity on whether the momentum balance we assume for this model (equating \emph{some form} of Eqs.~\ref{eq:pbub} \&~\ref{eq:psne}) is broken by this (perhaps modeling the momentum of the shock-front as receiving discrete `kicks' works well enough, \eg see \citealt{Seifried2018}).  Partly this is due to the fact that we aren't explicitly modeling the cooling in some continuous sense, and thus avoid the issues that occur in simulations when one injects `fractional' SNe energy/momentum into the ISM (see discussions of the `overcooling problem' in, \eg \citealt{Simpson2015, Hu2017, Su2018}).

For the most part, since the GMC mass function is fairly shallow, most star formation occurs in the most massive GMCs \citep[\eg half of the molecular gas in the Milky Way is contained in clouds with mass greater than $M_{\rm cloud} = 8.4\times10^5$ M$_\odot$, ][]{Miville-Deschenes2016}, and these will form star clusters with well-sampled IMFs.  And so it follows that those massive GMCs are predominantly responsible for producing the feedback that regulates disks.  We thus believe that it is reasonable to argue for this model that we can assume most feedback occurs in star clusters with a well-sampled IMF, falling on the star cluster formation efficiency scalings of \citet{Grudic2018}.
    
This model neglects \emph{triggered} star formation \citep[which is semantically fraught,][]{Dale2015}, assuming only that star cluster formation follows from some disk-scale (hydrodynamic) instability occurring on roughly an orbital dynamical time that results in fragmentation and collapse on a Toomre-length, and not from processes like cloud-cloud collisions. Simulation work (at higher redshift) by \citet{Ma2020} has shown that superbubbles themselves are capable of triggering star formation events in the dense shock region itself as it propagates through a dense turbulent patch of a galaxy.  This suggests that this model may \emph{underestimate} both the average star formation rate per star formation event and the number of SNe that will occur inside of a given superbubble, given that any triggering events (occurring at rates greater than once per orbital dynamical time) ought to increase star formation rates across galaxy disks \citep{Jeffreson2018}.  \rev{Further, star clusters that form near the edge of a previous superbubble may themselves be more likely to break out of the ISM (though the scalings used by our model for star cluster formation may not hold in such triggered star formation events).}

\subsection{Superbubbles and Galaxy Evolution/Disk Settling}\label{sec:galevolve}
Though much of the discussion surrounding this model so far has been framed in terms of superbubble evolution in \emph{disk} environments, we argue that none of our assumptions are strongly dependent on the thin disk geometry, \ie this model generally ought to be applicable to higher-redshift, less disky galaxies too, as long as $\tilde Q_{\rm gas} \sim 1$ and $v_{rot} \gtrsim \sigma$. Addressing these head-on: (1) the star cluster formation model of \citet{Grudic2018} is \emph{almost} entirely agnostic to redshift and disk environment, as they validated the model to relatively low metallicities ($\sim$10$^{-2} Z_\odot$) and make no assumption regarding the wider geometry or context of their star forming clouds; (2) the connection between scale height, velocity dispersion, and $\Omega$, \ie $H = \sigma/\Omega$, is generally true of gas on circular orbits (provided it is not \emph{strongly} self-bound, which would introduce a $1/\sqrt{G\rho}$ term) as an argument of simple harmonic motion about the orbit mid-plane; (3) though Toomre's $Q$ was originally derived for a thin, axisymmetric disk \citep{Toomre1964}, \emph{some quantity} with an identical scaling (admittedly with differing order unity pre-factors) is inescapable when deriving the condition for the balance between tidal forces (shear), turbulent/thermal pressure support, and gravity.  We need not only use $\tilde Q_{\rm gas}$ in \emph{disk} environments-- any scale-height cloud will do.

Both of the primary physical parameters of this model, the local gas fraction and dynamical time, generally evolve with redshift: gas fractions and dynamical times \emph{on-average} fall with time ($\tilde f_{\rm g} \downarrow$ and $\Omega \uparrow$, albeit with the latter only  weakly evolving), as galaxies deplete their gas reservoirs following gas peak accretion around $z \sim 2$ and as their rotation curves rise with halo assembly ($t_{\rm dyn} \approx 1/\Omega = R/v_c$).  And so, we expect that the distribution of independent star-forming regions in galaxies in $\tilde f_{\rm gas}$--$\Omega$ space to evolve towards lower $\tilde f_{\rm gas}$ and higher $\Omega$, respectively.  Though, since most galaxies are believed to have inside-out star formation profiles, the edges of star-forming galaxies should generally always inhabit the high-gas-fraction, long-dynamical-time (low $\Omega$) region of this parameter space, \ie near \textbf{CBO case}-conditions.  Extreme galactic outskirts also happen to be the setting where our assumptions of (1) a turbulently, as opposed to thermally, supported ISM and (2) an adequately sampled IMF break down (see \S~\ref{sec:modelfail}).  Evolutionarily speaking, it is then the centers of galaxies that are most interesting to consider for this model.  Necessarily, the central cores of galaxies have the shortest dynamical times and must fall from on-average having high local gas fractions at high redshift to low gas fractions at late times. 

The evolution of the central regions of galaxies, falling from high to low gas fractions, may then be tied directly to the ability of the central nuclei to host SN-driven outflows.  With superbubbles driving outflows, and disrupting gas orbits more generally in the galaxy protodisks, this transition from a galactic center with \textbf{PBO} to \textbf{PS case} outcomes could be a mechanism that allows the gas to settle into thin disks (as the turbulence decays over a few dynamical times with no major disruption), and a more general settling-down of star formation rate variability (\ie `burstiness') with galaxy (proto)disks quickly achieving the `classic' KS equilibrium (\eg the models of \citealt{Ostriker2011}, \citealt{Faucher-Giguere2013} or \citealt{Hayward2017}; seen in recent simulations by \citealt{Martizzi2020}). This is related to the central starburst--feedback instability studied by \citet{Torrey2017}, where galactic centers lose the ability to regulate themselves with feedback when the dynamical timescale becomes shorter than the stellar evolutionary timescale.  There is indeed some preliminary evidence of this being the case in the formation histories of the Milky Way-mass FIRE-2 galaxies (Gurvich, private correspondence), with it being tentatively seen that the transition from break-outs to contained superbubbles in the central parts of protodisks preceding settling of SFRs and gas orbit circularization.  Future large surveys at intermediate redshift, $z \approx 1-2$, where most disk formation appears to happen for $L^\star$ galaxies, with the ability to detect outflows and ascertain galaxy shapes (\eg the Roman/WFIRST High Latitude Spectroscopic Survey, HLSS; \citealt{Green2012}) may provide the necessary evidence to determine if this mechanism is indeed a driving cause of disk settling/formation or necessary for SFR equilibria.

\section{Summary \& Conclusions}\label{sec:conclusion}

In this paper, we presented a simple model for the clustering of SNe feedback from star clusters in disk environments and its effects on the ISM, specifically following the ability of superbubbles of SNe remnants, created by spatially and temporally overlapping SNe from a central star cluster, to break out of the gas disk of a galaxy.  The central result, similar to that of \citet{Hayward2017}, is that this feedback can both drive turbulence within the disks and outflows/fountains out of the disk into the near CGM.  First, the SNe inject momentum into the dense gas, driving turbulence, in a momentum-conserving shell.  If breakout then occurs (dependent on local disk conditions, \ie gas fractions and dynamical times), subsequent SNe do not inject their momentum into the dense gas but rather into galactic outflows and fountains.  Thus, there is a trade off, dependent on the occurrence and relative timing of superbubble breakout, between the amount of feedback injected into the dense gas of the ISM and that put into the CGM.  For astrophysicists interested in star formation in galaxies, one might consider the ``effective'' strength of stellar feedback, even though the actual evolution and feedback from the individual massive stars does not change.

Several key takeaways from this model include the following:
\begin{itemize}

\item Superbubble break-outs (and thus galactic outflows/fountains) are \emph{almost always powered}, in the sense that SNe are still occurring when the remnant reaches the edge of the disk (see \S~\ref{sec:case0III} \& \ref{sec:case12}).  If the remnant fails to reach the edge of the disk \emph{before} SNe stop occurring, it nearly always stalls and fragments within the disk, whereupon a potentially large fraction of the feedback momentum is deposited locally on a length scale below that of the disk scale height. 

\item In low gas fraction disks, where superbubbles fail to break out (`\textbf{PS case}') and so inject the majority of their feedback momentum at the stall/fragmentation scale, the driving scale of turbulence is smaller than the gas disk scale height and may be detectable in the power spectrum of its turbulent cascade.  This may also act as a regulator of the gas disk scale height (or star-forming clump size, in non-disk environments), with disks/clumps evolving towards fragmenting near the scale height.

\item Under the conditions of local star-forming galaxies, this model suggests that should break-out occur (`\textbf{PBO case}') upwards of 60\% of the feedback momentum will be channeled into outflows/fountains and \emph{not} drive turbulence in the gas disk. This could result in an apparent steepening (or offset, in the case of high-$z$ starbursts) of the Kennicutt-Schmidt relation as the effective strength of feedback falls (see \S~\ref{sec:KS}).

\item This model predicts that nearly all superbubble-driven outflows should include a cold component (launched initially at the shock break-out velocity) with abundance patterns similar to the ambient ISM and a hot component (corresponding to nearly all of the outflow energy loading) with abundances closely resembling SN direct ejecta.  Furthermore, it also predicts that the cold component would be moving relatively slowly, with a maximal speed a few times the local velocity dispersion (see \S~\ref{sec:outflows}).

\end{itemize}

The interpretations of this model, which is notably built on the consequences of overlapping post-Sedov-Taylor SN remnants in the momentum conserving `snowplow' phase, are most applicable to understanding why and how star formation feedback either couples mostly into the ISM or the near-CGM, and for calibrating feedback models in \rev{large-box cosmological} simulations which do not resolve the full evolution of individual SN remnants. 


\begin{acknowledgments} 
MEO is grateful for the encouragement of his late father, SRO, in studying astrophysics.
We thank Alex Gurvich, Lee Armus, and Phil Hopkins for conversations relating this model to disk formation at intermediate redshifts, and connections with spatially resolved observations and superwinds.  \rev{We also would like to thank the anonymous reviewer for their comments that greatly improved the manuscript.}
MEO was supported by the National Science Foundation Graduate Research Fellowship under Grant No. 1144469.
The Flatiron Institute is supported by the Simons Foundation.  We thank Lucy Reading-Ikkanda/Simons Foundation for assistance with developing Figure \ref{fig:cartoon}. 
This research has made use of NASA's Astrophysics Data System.
B.B is grateful for support from the Packard Fellowship and Sloan Fellowship.
\end{acknowledgments}




\bibliographystyle{aasjournal} 
\bibliography{library,alt_bibs} 



\appendix

\section{Validating SN Rate Model with Starburst99}\label{appendix:SNrate}

\begin{figure}
\centering
	\includegraphics[width=0.97\columnwidth]{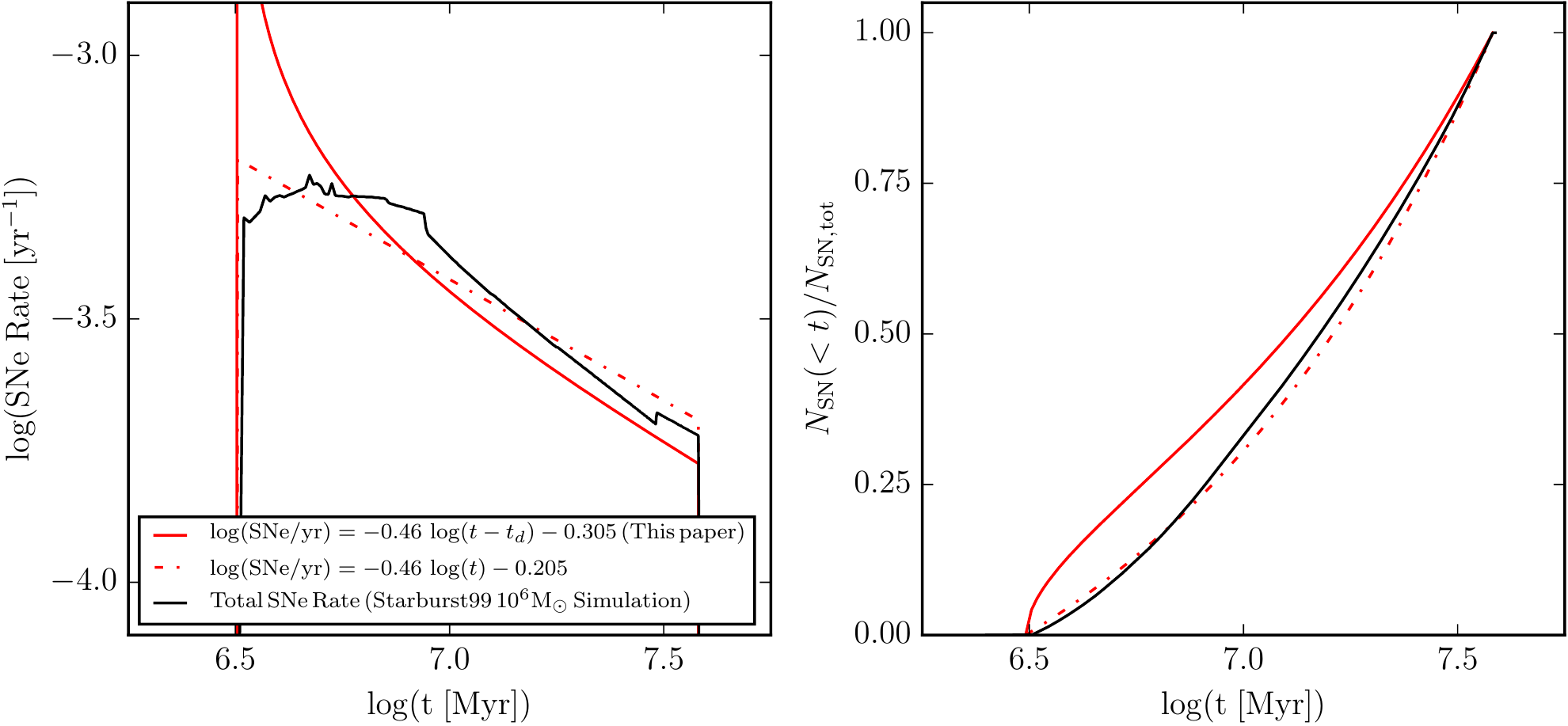}
    \caption{Comparing the SN delay time distribution model used in this paper (solid red line), $dN_{\rm SN}/dt \propto t^{-0.46}$ with a normalization of 1 SN/100 M$_\odot$ and SNe occurrence duration of $t_{\rm SNe} = 35$ Myr, against a Starburst99 \citep{Leitherer1999, Leitherer2014} simulation of a 10$^6$ \msolt cluster (instantaneously formed, solid black line). Solid versus dashed red lines show the difference in SN rate when incorporating a non-zero delay time (\ie $t^{-\alpha}$ vs. $(t+t_d)^{-\alpha}$), neglecting the delay time results in a divergence at the start of the SN period.  Our model is in good agreement with the stellar population synthesis/evolutionary model, though the integrated normalization ($N_{\rm SN}$ per stellar mass formed) of the Starburst99 model is $\approx 5\%$ higher over the duration of the feedback event than our model.  In addition to simple theory arguments relating to the high-mass IMF slope and stellar lifetimes (see \S~\ref{sec:remnantevolution}), and a SN progenitor mass cutoff (see \S~\ref{appendix:parameters}), this is a relatively strong justification for our choices of SN delay time distribution slope, duration of the occurrence SNe, and overall normalization of SNe per stellar mass formed.  Neglecting the $t_d$ term does result in a divergence in rate at $t=0$ for the $t-t_d$ model (this paper), but only slightly more $\lesssim 10\%$ \emph{cumulative} SNe at early times ($t \sim t_d$), and does not \emph{qualitatively} change the results.}
    \label{fig:SB99_SNrate}
\end{figure}

Here we empirically justify our choice of a $\propto t^{-0.46}$ power law, and effectively no delay time ($t_d = 0$), for modelling the delay time distribution of core-collapse SNe following the formation of a star cluster. In addition to theoretical arguments suggesting a power law with this exponent (see Appendix~A of \citealt{Orr2019}), we compare with outputs from {\scriptsize STARBURST99} \citep{Leitherer1999, Leitherer2014}.  Figure~\ref{fig:SB99_SNrate} shows the rate of supernovae, from a $10^6 M_\odot$ star cluster, and the number of supernovae that have already detonated over the total supernova number, for both outputs from {\scriptsize STARBURST99} (assuming their standard model parameters, as of writing, at \url{https://www.stsci.edu/science/starburst99/docs/parameters.html}) and two $\propto t^{-0.46}$ power law rates with the same overall supernovae number normalization (1 SN/100 M$_\odot$).  To solve for closed forms in this model, we neglect the delay time term in the rate of supernova (\ie $dN_{\rm SN}/dt \propto (t+t_d)^{-\alpha} \rightarrow t^{-\alpha}$), which results in a divergence of the supernova rate at $t=0$.  However, since the \emph{cumulative} momentum from supernovae that have occurred is what we consider for the evolution of the superbubble, this divergence does not qualitatively alter the resulting closed form solutions for, \eg superbubble radius or shock-front velocity.  The relative amount of supernovae occurring \emph{earlier} than stellar lifetimes/IMF slope would suggest, or predicted by {\scriptsize STARBURST99} is at most $\approx 10\%$.  And again, this difference is greatest for $t \lesssim t_d \approx 3.5$~Myr, reducing to insignificance for times $> 10$~Myr.  A $dN_{\rm SN}/dt \propto t^{-0.46}$ form very closely matches the {\scriptsize STARBURST99} population synthesis model, and neglecting the $t_d \approx 3.5$~Myr delay time does not dramatically alter the resulting \emph{cumulative} SN delay time distribution.

\section{Case Boundary Dependence on Physical Parameters}\label{appendix:parameters}

\begin{figure*}
\centering
	\includegraphics[width=\textwidth]{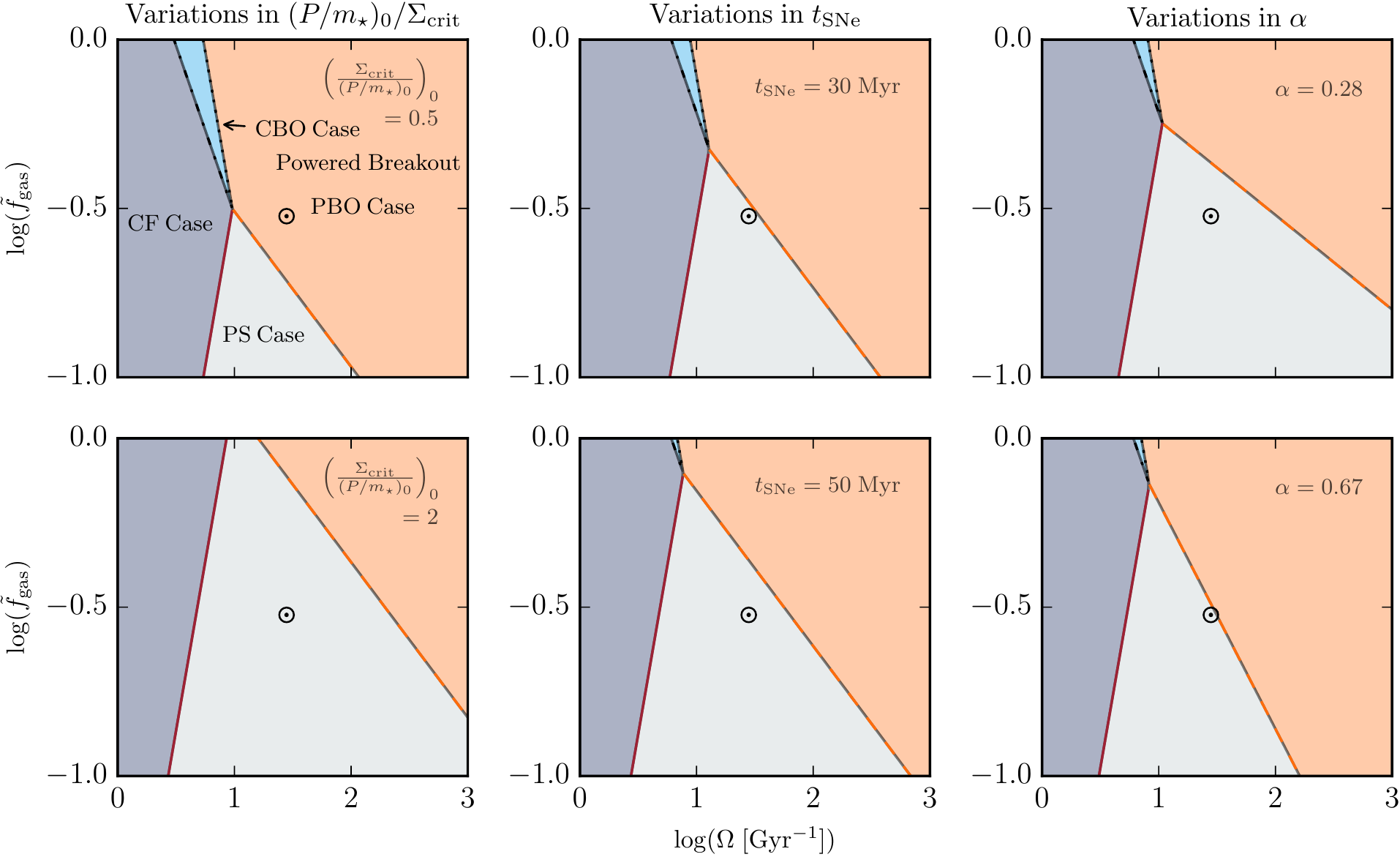}
    \caption{Case boundaries with variations in the assumed physical parameters for the superbubble model, in the style of Figure~\ref{fig:case}, in gas fraction--dynamical time phase space, following Eqs.~\ref{eq:case0III}, \ref{eq:tbotsn}, \ref{eq:case23} \& \ref{eq:case12}.  \textbf{Left panels:} varying the ratio of $\Sigma_{\rm crit}/(P/m_\star)_0$, the ratio of the strength of feedback to the gas surface density at which star cluster formation efficiency saturates.  Stronger feedback/more efficient cluster formation (weaker FB/less efficient SF) increases (decreases) the likelihood of breakout, powered or not.  At some point, with weak feedback, or inefficient star cluster formation, the \textbf{CBO case} becomes impossible. 
    \textbf{Center panels:} varying the duration of feedback injection $t_{\rm SNe}$. Shortening (lengthening) the duration of feedback is nearly equivalent to increasing (decreasing) its intensity/strength. 
    \textbf{Right panels:} changing the power-law slope of the SN time delay distribution $\alpha$. Increasing $\alpha$ (steeper slopes) makes powered break-out more likely.  On the other hand, as $\alpha \rightarrow 0$, the boundary between \textbf{PBO} \& \textbf{PS cases} tends towards a constant $\tilde f_g$.}
    \label{fig:cases_var}
\end{figure*}

In exploring the parameter space of superbubble outcomes, our model only has three remaining free parameters, once $\tilde Q_{\rm gas}=1$ is assumed along with the cluster formation efficiency model of \citet{Grudic2018}: the slope of the SNe time delay distribution $\alpha$, the duration of SNe momentum injection (the time from the first until the last Type-II SNe to occur in a young star cluster) $t_{\rm SNe}$, and the ratio of the gas surface density at which star cluster formation efficiency saturates to the fiducial strength of feedback from an individual supernova, \emph{i.e.}, $\Sigma_{\rm crit}/(P/m_\star)_0$. Figure~\ref{fig:cases_var} demonstrates how variations in each of these three effective parameters affect the case boundaries in this superbubble model.  Though the first two parameters, $\alpha$ and $t_{\rm SNe}$, being related both to the stellar IMF and stellar lifetimes, are not truly independent, we nonetheless treat them as separable.  

Reasonable uncertainties on the lower mass limit of Type-II progenitors at $8 \pm 1$ M$_\odot$ \citep{Smartt2009}, yield a range for $t_{\rm SNe}$ of 30-50 Myr.  Similarly, assuming that there exists an error of power-law slope in each the stellar IMF and mass to luminosity ratios of $\pm 0.2$, from $-2.35$ and $3.5$ \citep{Kroupa2002, Boehm-Vitense1992}, respectively, results in a range (assuming the maximum conspiracy between the two errors in either direction) of $\alpha = 0.28-0.67$.  It is not thought that the uncertainties on either is that large in the local universe \citep{Guszejnov2018, Guszejnov2019}.  Lastly, the fiducial momentum injected by a single supernova remnant into the ISM, \ie the `strength of feedback' $(P/m_\star)_0$ in terms of momentum injected by mass of stars formed, has uncertainties on the factor of two level \citep{Martizzi2015}. 

Variations in each of the three physical parameters predominantly move the critical point in $\tilde f_g$--$\Omega$ space where all four cases intersect, and qualitatively the outcomes expected in disk galaxies at late times do not change: coasting outcomes (\textbf{CBO/CF cases}) are unlikely, though shorter duration, more intense feedback (shorter $t_{\rm SNe}$, higher $(P/m_\star)_0$) increase the likelihood.  Weaker feedback, or a more gradually falling SN time delay distribution, both increase the difficulty of achieving breakout/driving outflows.

\section{A Uniform Supernova Delay Time Distribution ($\alpha = 0$)}\label{appendix:alpha0}
\begin{figure}
\centering
	\includegraphics[width=0.5\columnwidth]{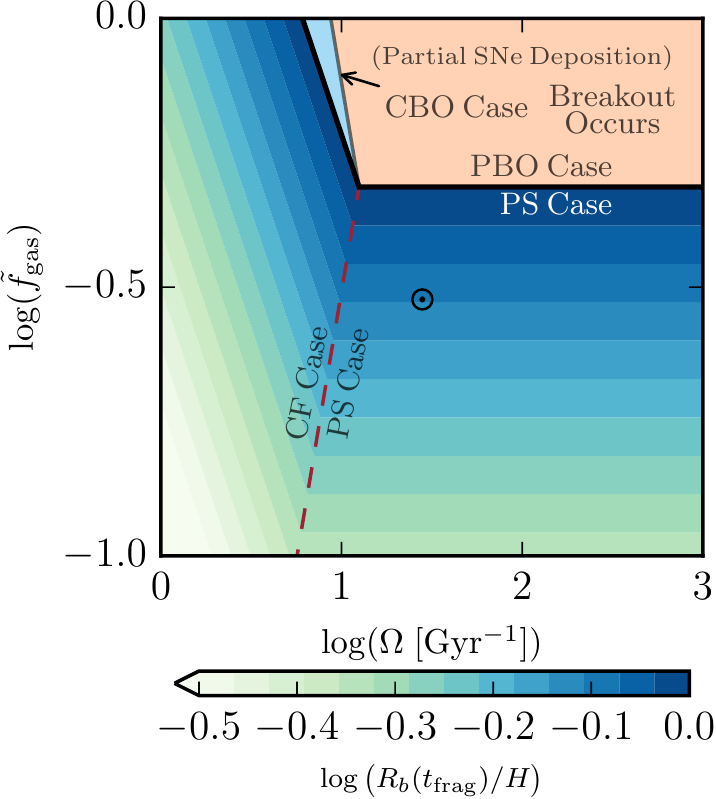}
    \caption{In the style of Figure~\ref{fig:fragscale}, fragmentation radius as a fraction of disk scale height of superbubbles that fail to break out of the disk (\textbf{CF/PS cases}), in gas fraction--dynamical time phase space, following Eqs.~\ref{eq:fragscale-case2} \& \ref{eq:fragscale-case3}, in the case of $\alpha=0$. The most dramatic difference with $\alpha =0 $ being that the boundary between \textbf{PBO} and \textbf{PS cases} becomes a constant line of $\tilde f_g$ above $\Omega = 1/2t_{\rm SNe}$ (see \S~\ref{sec:critpt}).}
    \label{fig:fragscale_alpha0}
\end{figure}

\begin{figure}
\centering
	\includegraphics[width=0.5\columnwidth]{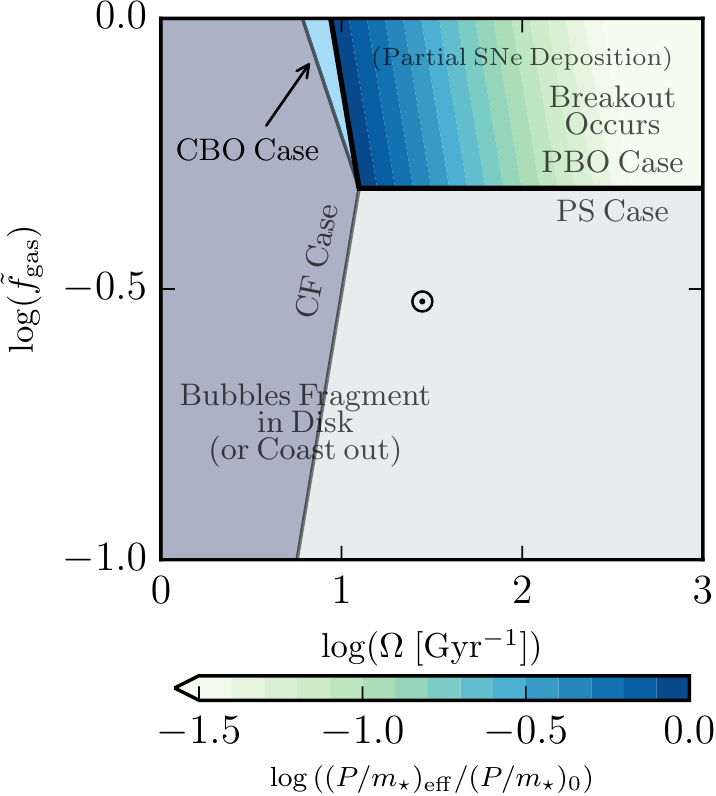}
    \caption{In the style of Figure~\ref{fig:effPMS}, ratio of `effective' to fiducial feedback strength of superbubbles that successfully break out of the disk \emph{before} $t_{\rm SNe}$ in gas fraction--dynamical time phase space, following Eq.~\ref{eq:pmseff}, when $\alpha=0$. The most dramatic difference with $\alpha =0 $ being that the boundary between \textbf{PBO} and \textbf{PS cases} becomes a constant line of $\tilde f_g$ above $\Omega = 1/2t_{\rm SNe}$ (see \S~\ref{sec:critpt}). The discontinuity of $(P/m_\star)_{\rm eff}$ along the \textbf{PBO/PS case} boundary remains.}
    \label{fig:effPMS_alpha0}
\end{figure}
As discussed in \S~\ref{sec:case0III}, we can imagine a universe where various physics have conspired such that the high-end slope of the IMF and massive stellar lifetimes result in a uniform time delay distribution of Type II SNe (if we assume that massive stellar lifetimes scale as $t_\star \propto M_\star/L_\star$, then this requires that the slope of the IMF be inverse to that of the mass to luminosity ratio, \emph{i.e.}, $dN/dM_\star \propto M^{-\beta}_\star$ and $L_\star \propto M_\star^\beta$, which is not currently believed to hold in the local universe).  In this case, $\alpha =0$, and the boundary between powered break-out (\textbf{PBO case}) and stall (\textbf{PS case}) becomes a constant $\tilde f_{g, \alpha=0} = 2\sqrt{2}\pi G \Sigma_{\rm crit} t_{\rm SNe}/3 (P/m_\star)_0\approx 0.5$ when including our erstwhile fiducial values.  The boundaries between \textbf{PBO} and \textbf{CBO cases} as well as \textbf{CF} and \textbf{PS cases} are only (weakly) affected in their normalization, not their slopes.  Figures~\ref{fig:fragscale_alpha0} \& \ref{fig:effPMS_alpha0} show both these changes to the case boundaries, as well as the bubble fragmentation scale (Eqs.~\ref{eq:fragscale-case3} \&~\ref{eq:fragscale-case2}) and effective strength of feedback (Eq.~\ref{eq:pmseff}), when $\alpha=0$. Broadly speaking, it remains unlikely that in such circumstances we should see coasting outcomes (\textbf{CBO/CF cases}), and outflows/break-outs becomes harder to achieve in disk galaxies at late times.


\label{lastpage}
\end{document}